\documentclass[%
reprint,
superscriptaddress,
 amsmath,amssymb,
 aps,
]{revtex4-1}

\usepackage[utf8]{inputenc}
\usepackage{graphicx}
\usepackage{dcolumn}
\usepackage{bm}
\usepackage{mathrsfs}
\usepackage{ytableau}

\usepackage{hyperref}
\hypersetup{
  colorlinks = true,
  urlcolor = blue,
  linkcolor = blue,
  citecolor = green,
  filecolor = magenta,
}

\newcommand{\ud}{\mathrm{d}}
\newcommand{\pd}{\partial}

\newcommand{\x}{\hat{\vartheta}}
\newcommand{\y}{\hat{\varphi}}
\newcommand{\ho}{\hat{\Omega}}
\newcommand{\hb}[1]{\hat{\boldsymbol{#1}}}
\newcommand{\bb}[1]{\bar{\boldsymbol{#1}}}
\newcommand{\tb}[1]{\tilde{\boldsymbol{#1}}}
\newcommand{\cb}[1]{\breve{\boldsymbol{#1}}}

\newcommand{\brb}[1]{\mathring{\boldsymbol{#1}}}
\newcommand{\sfe}{\mathsf{e}}
\newcommand{\fre}{\mathfrak{e}}

\newcommand{\stvalue}{10^{-2}}
\newcommand{\mathringsv}{10^{-3}}
\newcommand{\ovalueI}{\pi/2}
\newcommand{\ovalueII}{\pi/4}

\begin{document}


\title{Nontensorial gravitational wave polarizations from the tensorial degrees of freedom: \\I. Linearized Lorentz-violating theory of gravity with $\hb s$ tensor}

\author{Shaoqi Hou}
\email{hou.shaoqi@whu.edu.cn}
\affiliation{School of Physics and Technology, Wuhan University, Wuhan, Hubei 430072, China}
\author{Xi-Long Fan}
\email{xilong.fan@whu.edu.cn}
\affiliation{School of Physics and Technology, Wuhan University, Wuhan, Hubei 430072, China}
\author{Tao Zhu}
\email{zhut05@zjut.edu.cn; Corresponding author}
\affiliation{Institute for Theoretical Physics and  Cosmology, Zhejiang University of Technology, Hangzhou, Zhejiang 310032, China}
\author{Zong-Hong Zhu}
\email{zhuzh@whu.edu.cn}
\affiliation{School of Physics and Technology, Wuhan University, Wuhan, Hubei 430072, China}
\affiliation{Department of Astronomy, Beijing Normal University, Beijing 100875,  China}

\date{\today}

\begin{abstract}
General relativity predicts the existence of only two tensorial gravitational wave polarizations, while a generic metric theory of gravity can possess up to four additional polarizations, including two vectors and two scalar ones. 
These vector/scalar polarizations are in general generated by the intrinsic new vector/scalar degrees of freedom of the specific theories of gravity. 
In this paper, we show that, with the violation of the Lorentz symmetry in the framework of the standard model extension, the additional nontensorial polarizations can be directly excited by the two tensorial degrees of freedom.
We consider the diffeomorphism invariant standard model extension in the gravity sector with the Lorentz-violating coefficients $\hb s^{(d)\mu\rho\nu\sigma}$ of the even mass dimension $d\ge4$.
In addition to the extra polarizations induced by the tensor modes, the gravitational wave in this theory travels at a speed depending on the propagation direction, experiences dispersion if and only if $d\ge6$, and possesses neither velocity nor amplitude birefringence.
The excitement of the extra polarizations is also chiral.
The antenna pattern functions of interferometers due to such kinds of gravitational waves are generally linear combinations of those for all polarizations.
Detected by pulsar timing arrays and the Gaia satellite, the stochastic gravitational wave background in this model could induce couplings among cross correlations, of the redshifts of photons and the astrometric deflections of the positions of pulsars, for different polarizations.
These characteristics enable the use of interferometers, pulsar timing arrays, and Gaia missions to constrain this model.
\end{abstract}

\maketitle


\section{Introduction}\label{sec-intro}

The direct detection of gravitational waves (GWs) from the coalescence of compact binary systems by the LIGO/Virgo/KAGRA Collaboration (LVK) has offered new opportunities to explore fundamental building principles of Einstein's general relativity (GR), including the equivalence principle, parity and Lorentz invariance, the spacetime dimension, etc \cite{gw150914, gw170817, LIGOScientific:2018mvr, LIGOScientific:2020ibl, KAGRA:2021vkt}. 
One central prediction of GR is the existence of only two independent polarization modes, the tensorial plus and cross modes, which propagate at the speed of light with an amplitude damping rate inversely proportional to the luminosity distance of the GW source. 
Any violation of the fundamental principles of GR could lead to possible derivations from the above standard propagation properties of GWs and even generate extra polarization modes beyond the two tensorial ones. 
An essential observation from the generic metric theories of gravity is that they can allow for up to four additional polarizations: the vector-$x$ and vector-$y$ polarizations and the breathing and longitudinal scalar ones \cite{Eardley:1973br,Eardley:1974nw,Will:2014kxa}. 
The detection or absence of these extra polarizations can provide pivotal insights into the fundamental nature of gravity, potentially revealing deviations from GR and helping constrain or even refuting alternative theories.

It is folklore that whenever a modified theory of gravity provides extra degrees of freedom (DoF's), there are non-Einsteinian GW polarizations excited by the extra DoF's \cite{Will:2014kxa,Hou:2017bqj,Gong:2018ybk,Gong:2018cgj,Gong:2018vbo,Dong:2021jtd,Dong:2023bgt}.
Usually, the vectorial DoF's induce vector-$x$ and vector-$y$ polarizations, the scalar DoF's excite the longitudinal and breathing polarizations, and of course, the tensorial DoF's cause  $+$ and $\times$ polarizations.
These well-known results definitely hold for a whole plethora of theories of gravity, including general relativity (GR) \cite{mtw}, scalar-tensor theories \cite{Eardley:1973br,Eardley:1974nw,Liang:2017ahj,Hou:2017bqj}, vector-tensor theories \cite{Gong:2018cgj}, and candidates of quantum gravity \cite{Gong:2018vbo}, etc.
In this paper, we describe a mechanism that the nontensorial polarizations can be directly generated from the two tensorial DoF's,  due to the violation of the Lorentz invariance in the framework of the standard model extension (SME)

The SME \cite{Kostelecky:2003fs} is an effective field theory incorporating new terms responsible for the local Lorentz violation and CPT violation into the action of the standard model of elementary particles \cite{Weinberg:1995mt} and Einstein-Hilbert action \cite{mtw}.
These new terms are Lorentz-violating operators with coefficients carrying Lorentz indices.
These coefficients may arise from some symmetry-breaking mechanism, either spontaneous or explicit.
They preferably define a certain local Lorentz frame, violating the local Lorentz invariance. 
In the framework of SME, one should clearly distinguish two kinds of local Lorentz transformations, the observer and the particle transformations.
The former refers to the change in the local Lorentz frame. 
The latter transforms particles or local fields but leaves the Lorentz-violating operators unchanged.
In the current work, we consider the case of the observer diffeomorphism invariant theory, and the coefficients of the Lorentz-violating operators are taken to be constant in the flat spacetime limit. 
In the following, we will simply use ``diffeomorphism'' to refer to ``observer diffeomorphism'', unless specified.
To analyze the GW polarizations, in this work, the gravity sector is specifically considered, ignoring the parts of the action directly involving fields of elementary particles.

In the gravity sector of the SME, the Lorentz-violating terms can be grouped according to their mass dimensions $d$.
To maintain the diffeomorphism invariance, it is required that $d\ge4$ \cite{Kostelecky:2017zob}.
Expanded around the flat spacetime background, these operators take the following forms \cite{Kostelecky:2016kfm},
\begin{gather}
  \label{eq-def-lv-d}
  \mathcal L_d=\frac{1}{4}h_{\mu\nu}\hat{\boldsymbol{\mathcal K}}^{(d)\mu\nu\rho\sigma}h_{\rho\sigma},\\
  \hat{\boldsymbol{\mathcal K}}^{(d)\mu\nu\rho\sigma}=\boldsymbol{\mathcal K}^{(d)\mu\nu\rho\sigma\alpha_1\cdots\alpha_{d-2}}\pd_{\alpha_1\cdots\alpha_{d-2}},
\end{gather}
where $h_{\mu\nu}=g_{\mu\nu}-\eta_{\mu\nu}$, and $\pd_{\alpha_1\cdots\alpha_{d-2}}\equiv\pd_{\alpha_1}\cdots\pd_{\alpha_{d-2}}$.
A general Lorentz-violating operator is the sum of $\mathcal L_d$ over $d\,(\ge4)$.
According to the symmetries of the coefficient $\boldsymbol{\mathcal K}^{(d)\mu\nu\rho\sigma\alpha_1\cdots\alpha_{d-2}}$ under the permutations of its indices, it can be split into three pieces \cite{Kostelecky:2016kfm}, 
\begin{equation}
  \label{eq-split-K}
  \hat{\boldsymbol{\mathcal K}}^{(d)\mu\nu\rho\sigma}=\hat{\boldsymbol s}^{(d)\mu\rho\nu\sigma}+\hat{\boldsymbol{q}}^{(d)\mu\rho\nu\sigma}+\hat{\boldsymbol{k}}^{(d)\mu\nu\rho\sigma},
\end{equation}
each belonging to different conjugacy classes of the permutation group on the $(d+2)$ indices $\mu\nu\rho\sigma\alpha_1\cdots\alpha_{d-2}$ \cite{Georgi:2000vve}.
$\hat{\boldsymbol s}^{(d)\mu\rho\nu\sigma}$ and $\hat{\boldsymbol{k}}^{(d)\mu\nu\rho\sigma}$ are CPT even, and $\hat{\boldsymbol{q}}^{(d)\mu\rho\nu\sigma}$ is CPT odd.
Their impacts on the properties of $h_{\mu\nu}$, and in particular, the GW polarizations and propagation, are quite different.
The amplitude birefringence can be induced by $\hb{q}^{(d)\mu\rho\nu\sigma}$, while the velocity birefringence is caused by $\hb{q}^{(d)\mu\rho\nu\sigma}$ and $\hb{k}^{(d)\mu\nu\rho\sigma}$.
The tensor $\hb{s}^{(d)\mu\rho\nu\sigma}$ excites neither of the birefringence effects, but like $\hb q^{(d)\mu\rho\nu\sigma}$ and $\hb k^{(d)\mu\nu\rho\sigma}$, leads to modified GW speed, and anisotropic propagation of the GW.
In this work, let us study the effect of $\hat{\boldsymbol s}^{(d)\mu\rho\nu\sigma}$ on GW polarizations first.

As shown in the following discussion, the Lorentz-violating operators $\hb s^{(d)\mu\rho\nu\sigma}$ couple the tensor, vector and scalar modes of GWs together in such a way that all of the vector and scalar modes are excited by the tensor modes.
So there are only two tensorial DoF's.
They generally propagate at a speed different from the speed of light in vacuum (taken to be 1), and the dispersion occurs for $d>4$.
The speed also depends on the propagation direction.
Neither velocity nor amplitude birefringence takes place.
Since the vector and scalar polarizations are excited by the tensor DoF's, the antenna pattern functions of the interferometers for $+$ and $\times$ polarizations are different from the familiar ones \cite{Nishizawa2009,Isi:2017fbj}.
Now, they are linear combinations of the individual antenna pattern functions of various polarizations, as if they correspond to intrinsic DoF's.
In principle, one could use interferometers \cite{Harry:2010zz,TheLIGOScientific:2014jea,TheVirgo:2014hva,Somiya:2011np,Aso:2013eba,Audley:2017drz,Luo:2015ght,Taiji2017,Kawamura:2011zz} to detect the modified antenna pattern functions and constrain SME.
The stochastic GW background (SGWB) in this theory could also be detected by pulsar timing arrays (PTAs) \cite{Hobbs:2009yy,Kramer:2013kea,McLaughlin:2013ira,Hobbs:2013aka,2016ASPC..502...19L,Nobleson:2021ngl,Spiewak:2022btk} and Gaia mission \cite{Book:2010pf,2016AA...595A...1G,Moore:2017ity,Klioner:2017asb,Mihaylov:2018uqm,OBeirne:2018slh}. 
The motion of the photon coming from the pulsar or any star could be affected by the SGWB, leading to the changes in the measured frequency and propagation direction.
The change in the frequency of the photon is basically the redshift, and the change in the propagation direction causes the apparent position of the star to be altered, namely astrometric deflection.
One can form three basic types of correlation functions: redshift-redshift, astrometric-astrometric, and redshift-astrometric correlations.
The redshift-redshift correlation function is basically the cross-correlation function measured by PTAs \cite{Lee2008ptac}.
The astrometric-astrometric and redshift-astrometric correlations can be monitored by Gaia mission \cite{2016AA...595A...1G}.
At least, some of them are expected to be different from the ones predicted by GR, such as the Hellings-Downs (HD) relation for the correlation between redshifts \cite{Hellings:1983fr}, and the standard relations for the astrometric deflections \cite{Book:2010pf} and redshift-astrometric \cite{Mihaylov:2018uqm}.
So one may use PTAs or Gaia mission to constrain this theory, too.
In this work, we consider all of these correlations, as the SME contains a lot of Lorentz-violating coefficients, and different correlations are sensitive to different coefficients.
Interferometers usually bound a theory in the higher frequency regions ($\sim 10^{-4} - 10^4$ Hz), while PTAs and Gaia mission provide data in the lower frequency ranges ($\sim 10^{-10} - 10^{-6}$ Hz).

The observations of GW170817 and GW170817A have placed a very strong bound on the tensor GW speed \cite{TheLIGOScientific:2017qsa,Goldstein:2017mmi,Savchenko:2017ffs}.
Many modified theories of gravity are thus highly constrained, including Horndeski theory \cite{Baker:2017hug,Creminelli:2017sry,Sakstein:2017xjx,Ezquiaga:2017ekz,Langlois:2017dyl,Sakstein:2017xjx,Gong:2017kim}, Ho\v{r}ava theory \cite{Gumrukcuoglu:2017ijh,Gong:2018vbo}, and Einstein-\ae theor theory \cite{Oost:2018tcv} and generalized TeVeS theory \cite{Gong:2018cgj}.
Since the theory considered in this work generally predicts a speed different from 1, its coupling constants should also be severely restricted.
However, its speed actually is a function of the propagation direction. 
The speed bound effectively places the constraint on one of the coupling constants.
With signals of GW events in LVK catalogs, the Lorentz-violating effects on gravitational waveform of the two tensorial modes and their constraints  have been studied in a lot of works \cite{Kostelecky:2016kfm, Niu:2022yhr, Haegel:2022ymk, ONeal-Ault:2021uwu, Wang:2017igw, Zhao:2022pun, Gong:2023ffb, Wang:2021gqm, Shao:2020shv, Wang:2021ctl, Zhu:2023wci}.
Most recently, North American Nanohertz Observatory for Gravitational Waves (NANOGrav) \cite{NANOGrav:2023gor}, Parkes Pulsar Timing Array (PPTA) \cite{Reardon:2023gzh}, European Pulsar Timing Array (EPTA) and Indian Pulsar Timing Array (InPTA) \cite{EPTA:2023sfo,EPTA:2023fyk}, and  Chinese Pulsar Timing Array (CPTA) \cite{Xu:2023wog} announced the evidence for a stochastic signal that conforms to the HD relation.
This implies constraints on the modified theories of gravity.
One may use their data to constrain the SME. 
However, as a first step in studying the effects of Lorentz violation on the GW polarizations, we would like to focus on the theoretical aspects. 
Using observational data to constrain the parameters of this theory will be done in a follow-up work.

There are works that discovered the excitation of the extra polarizations by the tensor modes, e.g., Refs.~\cite{Mewes:2019dhj,Liang:2022hxd,Bailey:2023lzy}.
The first work focused on the effects of dispersion and birefringence on GWs in SME. 
It did not discuss the responses of PTAs and Gaia satellites, which are important results of the current work.
The work~\cite{Liang:2022hxd} studied the GW polarizations in a special case of SME, the bumblebee gravity \cite{Kostelecky:2003fs}, while this work is more general.
Finally, Ref.~\cite{Bailey:2023lzy} analyzed the propagation of the scalar, vector and tensor fields in the linearized Lorentz-violating theories with matter sources.
The Lorentz-violating operator is also of mass-dimension 4.
The partial solution for the tensor field also shows the presence of extra polarizations, excited by the $+$ and $\times$ modes.

This work is organized in the following way.
In Section~\ref{sec-sme}, SME will be reviewed very briefly, and its GW solution above the flat spacetime background will be solved for using the gauge-invariant formalism.
Section~\ref{sec-gw-pol} will be devoted to the discussion of the GW polarization content.
There, the antenna pattern functions for the physical DoF's will be computed for the ground-based interferometers.
The responses of PTAs and Gaia mission to the SGWB in SME will be computed in Sections~\ref{sec-pta} and \ref{sec-gaia}, respectively.
Finally, there will be a conclusion in Section~\ref{sec-con}.
In the following, the label $P$ will refer to $+$ and $\times$ polarizations, specifically.
The extra polarizations will be denoted by $P'=\text{x},\text{y},\text{b}$ and $\text{l}$.
The units has been chosen such that $G=c=\hbar=1$.
Some of the calculations have been done with \emph{xAct} \cite{xact}.

\section{Standard model extension}
\label{sec-sme}

The action for the linearized gravity sector of the SME is given by \cite{Kostelecky:2016kfm},
\begin{equation}
    \label{eq-s-act}
    \begin{split}
    S=&\frac{1}{4}\int\ud^4x\big(\epsilon^{\mu\rho\alpha\kappa}\epsilon^{\nu\sigma\beta\lambda}\eta_{\kappa\lambda}h_{\mu\nu}\partial_{\alpha\beta} h_{\rho\sigma}\\
    &+h_{\mu\nu}\hat{\boldsymbol s}^{(d)\mu\rho\nu\sigma}h_{\rho\sigma}\big),
    \end{split}
\end{equation}
where $h_{\mu\nu}=g_{\mu\nu}-\eta_{\mu\nu}$, $\epsilon_{\mu\nu\rho\sigma}$ is the volume element compatible with $\eta_{\mu\nu}$, and the Greek indices are raised and lowered by $\eta^{\mu\nu}$ and $\eta_{\mu\nu}$, respectively.
$d$ is the mass dimension, even, and at least 4.
Instead of studying the case of a single $d$, one could consider a general Lorentz-violating operator, i.e., a sum over all possible $d$'s. 
However, the treatment for any $d$ is basically the same, so it is sufficient to consider the case of a particular $d$.
For this reason, the superscript $(d)$ will be omitted to make the following expressions less cluttered.
The first term of Eq.~\eqref{eq-s-act} is just the linearized version for the Einstein-Hilbert action \cite{mtw}. 
In the second line of Eq.~\eqref{eq-s-act}, one has
\begin{equation}
  \label{eq-def-so}
\hat{\boldsymbol s}^{\mu\rho\nu\sigma}=s^{\mu\rho\alpha_1\nu\sigma\alpha_2\alpha_3\cdots\alpha_{d-2}}\partial_{\alpha_1\alpha_2\cdots\alpha_{d-2}}.
\end{equation}
The tensor $s^{\mu\rho\alpha_1\nu\sigma\alpha_2\alpha_3\cdots\alpha_{d-2}}$ satisfies certain symmetries under the permutations of its indices, which is specified by the following Young tableau \cite{Kostelecky:2016kfm}, 
\begin{center}
\ytableausetup{centertableaux,boxsize=2.2em}
\begin{ytableau}
\mu & \nu & \alpha_3 & \none[\cdots] & \alpha_{d-2}\\
\rho & \sigma \\
\alpha_1 & \alpha_2
\end{ytableau}.
\end{center}
This also implies that $\hat{\boldsymbol{s}}^{\mu\rho\nu\sigma}$ is invariant under the infinitesimal coordinate transformation, and 
\begin{equation}
  \label{eq-s-pro}
  \begin{split}
s^{\mu\rho\alpha_1\nu\sigma\alpha_2\alpha_3\cdots\alpha_{d-2}}=&s^{[\mu\rho\alpha_1][\nu\sigma\alpha_2]\alpha_3\cdots\alpha_{d-2}}\\
=&s^{\nu\sigma\alpha_2\mu\rho\alpha_1\alpha_3\cdots\alpha_{d-2}}.
  \end{split}
\end{equation}
Due to this symmetry property, define a new operator $\bb{s}^{\mu\nu}(=\bb s^{\nu\mu})$ such that \cite{Kostelecky:2016kfm},
\begin{equation}
  \label{eq-def-bbs-d}
  s^{\mu\rho\alpha_1\nu\sigma\alpha_2\alpha_3\cdots\alpha_{d-2}}\pd_{\alpha_3\cdots\alpha_{d-2}}=-\epsilon^{\mu\rho\alpha_1\kappa}\epsilon^{\nu\sigma\alpha_2\lambda}\bb s^{}_{\kappa\lambda}.
\end{equation}
Note that $\bb s^{\mu\nu}$ carries $(d-4)$ partial derivatives, i.e., $\bb s^{\mu\nu}\equiv s^{\mu\nu\alpha_3\cdots\alpha_{d-2}}\pd_{\alpha_3\cdots\alpha_{d-2}}$.
One may further decompose $\bb s^{\mu\nu}=\tb s^{\mu\nu}+\eta^{\mu\nu}\bb s/4$ with $\tb s^{\mu\nu}$ traceless.
The action becomes
\begin{equation*}
    \label{eq-s-act2}
    S=\frac{1}{4}\int\ud^4x\epsilon^{\mu\rho\alpha\kappa}\epsilon^{\nu\sigma\beta\lambda}\bigg[\Bigl(1-\frac{\bb s}{4}\Bigr)\eta_{\kappa\lambda}-\tb s_{\kappa\lambda}\bigg]h_{\mu\nu}\partial_{\alpha\beta} h_{\rho\sigma}.
\end{equation*}
Therefore, the trace $\bb s$ of $\bb s^{\mu\nu}$ modifies the Newton's constant.
Generally speaking, the effective ``Newton's constant'' depends on the the GW frequency, speed and direction of the GW propagation for $d\ge6$.
For $d=4$, the effective Newton's constant is truly a constant.
As shown below, $\bb s$ does not affect the GW polarizations.

The equations of motion (EoM's) are given by \cite{Gong:2023ffb},
\begin{equation}
    \label{eq-s-eom}
    \epsilon^{\mu\rho\alpha\kappa}\epsilon^{\nu\sigma\beta\lambda}\bigg[\Bigl(1-\frac{\bb s}{4}\Bigr)\eta_{\kappa\lambda}-\tb s_{\kappa\lambda}\bigg]\partial_{\alpha\beta} h_{\rho\sigma}=0.
\end{equation}
Since the action is gauge invariant, it is beneficial to use  the gauge-invariant variables to solve the EoM's \cite{Flanagan:2005yc}.
So one decomposes the components of $h_{\mu\nu}$ in the following way,
\begin{gather}
  h_{tt}=2\phi,\\
  h_{tj}=\zeta_j+\pd_j\chi,\\ 
  h_{jk}=h_{jk}^\text{TT}+\frac{H}{3}\delta_{jk}+\pd_{(j}\varepsilon_{k)}+\left(\pd_{jk}-\frac{1}{3}\delta_{jk}\nabla^2\right)\rho,
\end{gather}
with $\pd_j\zeta^j=\pd_j\varepsilon^j=\delta^{jk}h_{jk}^\text{TT}=0$ and $\pd^k h_{jk}^\text{TT}=0$. 
Therefore, $\zeta_j$ is the transverse part of $h_{tj}$, and $\chi$ defines the curl-free component of $h_{tj}$, satisfying $\nabla^2\chi=\pd^jh_{tj}$.
$h_{jk}^\text{TT}$ is the transverse-traceless part of $h_{jk}$.
It can be checked that $H=\delta^{jk}h_{jk}$, $\nabla^2\nabla^2\rho=\frac{3}{2}\pd^{jk}h_{jk}-\frac{1}{2}\nabla^2H$, and $\nabla^2\varepsilon_j=2\pd^kh_{jk}-\frac{2}{3}\pd_jH-\frac{4}{3}\pd_j\nabla^2\rho$ \cite{Flanagan:2005yc}.
These components ($\phi,\chi,H,\rho$, $\zeta_j,\varepsilon_j$, and $h_{jk}^\text{TT}$) are generally all functions of the space and time.
Under an infinitesimal coordinate transformation, $x^\mu\rightarrow x^\mu+\xi^\mu$, one has
\begin{equation}
  \label{eq-h-gtf}
  h_{\mu\nu}\rightarrow h_{\mu\nu}-\pd_\mu\xi_\nu-\pd_\nu\xi_\mu.
\end{equation}
One shall also write $\xi_\mu=(A,B_j+\pd_jC)$ with $\pd^jB_j=0$.
Then, one can check that
\begin{gather*}
  \phi\rightarrow\phi-\dot A,\quad
  \chi\rightarrow \chi-A-\dot C,\\
  H\rightarrow H-2\nabla^2C,\quad
  \rho\rightarrow\rho-2C,\\
  \zeta_j\rightarrow\zeta_j-\dot B_j,\quad
  \varepsilon_j\rightarrow\varepsilon_j-2B_j,\\
  h_{jk}^\text{TT}\rightarrow h_{jk}^\text{TT}.
\end{gather*}
So the gauge-invariant variables are 
\begin{gather}
\Phi=-\phi+\dot\chi-\frac{1}{2}\ddot\rho,\\ 
\Theta=\frac{1}{3}(H-\nabla^2\rho), \\
\Xi_j=\zeta_j-\frac{1}{2}\dot\varepsilon_j,\\
h_{jk}^\text{TT}.
\end{gather}
As shown above, there are two scalar DoF's, two vector DoF's, and two tensor DoF's, although not of all them are independent.
It is easy to show that Eq.~\eqref{eq-s-eom} can be expressed solely in terms of these gauge-invariant variables, but unfortunately, not decoupled, as exhibited below,
\begin{subequations}
  \label{eq-eom-gi}
\begin{gather}
  \nabla^2(2\Theta+\tb s^{jk}h_{jk}^\text{TT})=0,\\
  \nabla^2\Xi_j+2\pd_j\dot\Theta+\tb s_0{}^k\nabla^2h_{jk}^\text{TT}+2\tb s^{kl}\pd_{[j}\dot h_{k]l}^\text{TT}=0,\\
  \Box h_{jk}^\text{TT}+2\pd_{(j}\dot\Xi_{k)}+2\delta_{jk}\ddot\Theta+(\pd_{jk}-\delta_{jk}\nabla^2)(2\Phi+\Theta)\nonumber\\
  +\tb s_{00}\nabla^2h_{jk}^\text{TT}+2\tb s_0{}^l(\pd_{(j}\dot h_{k)l}^\text{TT}-\pd_lh_{jk}^\text{TT})\nonumber\\
  +\tb s^{il}(\pd_{jk}h_{il}^\text{TT}+\pd_{il}h_{jk}^\text{TT}-2\pd_{i(j}h_{k)l}^\text{TT})=0.
\end{gather}
\end{subequations}
where $T_{[jk]}=(T_{jk}-T_{kj})/2$ and $T_{(jk)}=(T_{jk}+T_{kj})/2$.
To obtain these equations, one has omitted the terms that would be of the second order in $\tb s^{\mu\nu}$, after these equations are solved, otherwise, these expressions would be tremendously complicated.
When $\tb s^{\mu\nu}=0$, one recovers the gauge invariant equations as in GR \cite{Flanagan:2005yc}, and the solutions are simple,
\begin{gather}
  \Phi_\text{gr}=\Theta_\text{gr}=0,\quad \Xi_{\text{gr},j}=0,\\
 \Box h_{\text{gr},jk}^\text{TT}=0.
\end{gather}

It is easier to solve Eq.~\eqref{eq-s-eom} up to the linear order in $\tb s^{\mu\nu}$ in the momentum space. 
Let us assume 
\begin{equation}
  \label{eq-h-four}
h_{\mu\nu}=A_{\mu\nu} e^{i\Omega\cdot x},
\end{equation}
where $A_{\mu\nu}$ is a constant amplitude, and $\Omega^\mu=(\omega,\Omega\hat \Omega)$ with $\hat \Omega$ the unit vector in the propagation direction.
So all the partial derivatives $\pd_\mu$ in Eq.~\eqref{eq-eom-gi} shall be replaced by $i\Omega_\mu$.
Similarly, $\pd_\mu$ appearing in the operator $\bb s^{\mu\nu}$ becomes $i\Omega_\mu$, that is,
\begin{equation*}
  \label{eq-st-f}
  \bb s^{\mu\nu}=\bar s^{\mu\nu\alpha_3\cdots\alpha_{d-2}}\pd_{\alpha_3\cdots\alpha_{d-2}}\rightarrow i^d\bar s^{\mu\nu\alpha_3\cdots\alpha_{d-2}}\Omega_{\alpha_3}\cdots\Omega_{\alpha_{d-2}}.
\end{equation*}
Since $d$ is even, the rightmost term is always real, and we will still use $\bb s^{\mu\nu}$ to represent it.
Up to the linear order in $\bb s^{\mu\nu}$, one can show that
\begin{subequations}
  \begin{gather}
    \left(\omega^2-\Omega^2-\Omega_\mu\Omega_\nu\tb s^{\mu\nu}\right)h_{jk}^\text{TT}=0,\label{eq-tt-gi}\\
  \Phi=\frac{1}{2}\Theta=-\frac{1}{4}\tb s^{jk}h_{jk}^\text{TT},\label{eq-sc-gi}\\
  \Xi_j=-\hat\Omega_\mu\tb s^{\mu k}h_{jk}^\text{TT},\label{eq-vc-gi}
  \end{gather}
\end{subequations}
after some tedious algebraic manipulations of Eq.~\eqref{eq-eom-gi}.
Here, $\hat\Omega_\mu\equiv(1,\hat\Omega)$.
By Eq.~\eqref{eq-tt-gi},
the dispersion relation should take the following form,
\begin{equation}
  \label{eq-s-disp}
  \omega=c_\text{gw}\Omega,\quad c_\text{gw}=1+\frac{1}{2}\hat\Omega_\mu\hat\Omega_\nu\tb s^{\mu\nu},
\end{equation}
in order that it admits nontrivial wave solutions.
Therefore, the two tensor DoF's propagate at the same speed, which is generally different from the speed  of light in vacuum. 
There is no velocity birefringence.
For $d=4$, $c_\text{gw}$ is independent of the GW frequency, so there is no dispersion, while for larger $d$'s, the dispersion does happen.
The effect of dispersion would cause the dephasing of the GW signal after it travels a very long distance from the source to the detector, and the dephasing can be used to constrain the theory, as done in Ref.~\cite{Gong:2023ffb} using GWTC-3 data by LIGO/Virgo/KAGRA \cite{KAGRA:2021vkt}.
$c_\text{gw}$ depends also on the propagation direction $\hat\Omega^\mu$ of the GW, indicating the anisotropic propagation of the GW.
Given the violation of the Lorentz symmetry, it is permissible that the GW speed can be greater than 1, when $\hat\Omega_\mu\hat\Omega_\nu\tb s^{\mu\nu}>0$, as required by the observation of the gravitational Cherenkov radiation \cite{Elliott:2005va}. 
If $\hat\Omega^\mu$ happens to be an eigenvector of $\tb s^{\mu\nu}$, the GW propagates at $1$.
One can parameterize $\tb s^{\mu\nu}$ in the following way \footnote{In Eq.~\eqref{eq-tis}, we place a breve on top of $\boldsymbol s$ to indicate that $\cb s^{jk}$ is the traceless part of $\tb s^{\mu\nu}$. The similar notation was also used previously, for example, $\breve{k}^{\mu\cdots}{}_{\nu\cdots}{}^{a\cdots}$ in Ref.~\cite{Kostelecky:2020hbb}. However, their meanings are different. $\breve{k}^{\mu\cdots}{}_{\nu\cdots}{}^{a\cdots}$ carries both spacetime indices and the internal Lorentzian indices.  In addition, in the flat spacetime background, the Lorentzian indices of $\breve{k}^{\mu\cdots}{}_{\nu\cdots}{}^{a\cdots}$ may be identified with the spacetime indices, but they can take values in $\{0,1,2,3\}$. In contrast, $\cb s^{jk}$ has only spatial indices.}, 
\begin{equation}
  \label{eq-tis}
\tb s^{\mu\nu}=\left(
    \begin{array}{cc}
        \tb s & \brb s^j\\
        \brb s^k & \cb s^{jk}+\tb s\delta^{jk}/3
    \end{array}
\right),
\end{equation}
where $\delta^{jk}\cb s_{jk}=0$.
So the GW speed is 
\begin{equation}
  \label{eq-s-gws}
c_\text{gw}=1-\brb s_j\hat\Omega^j+\cb s_{jk}\hat\Omega^j\hat\Omega^k+\frac{2}{3}\tb s.
\end{equation}
One may conclude that the presence of $\tb s$ violates the Lorentz boost invariance, while $\brb s^j$ and $\cb s^{jk}$ define several spatial directions, breaking the rotational symmetry.
In fact, it is the breaking down of the rotational symmetry that allows the coupling between the tensor, vector and scalar modes.
The observations of GW170817 and its electromagnetic counterpart, GW170817A, have placed a very strong constraint on the GW speed \cite{TheLIGOScientific:2017qsa,Monitor:2017mdv},
\begin{equation}
  \label{eq-con-cgw}
 -3\times10^{-15}\le c_\text{gw}-1\le7\times10^{-16}.
\end{equation}
If the theory of gravity could be incorporated in the framework of SME, $|\hat\Omega_\mu\hat\Omega_\nu\tb s^{\mu\nu}|\le10^{-15}$, roughly \cite{TheLIGOScientific:2017qsa,Monitor:2017mdv}.
Note that this condition highly bounds one component of $\tb s^{\mu\nu}$, as $\hat\Omega_\mu$ for GW170817 is fixed, in principle.
The rest components are less constrained.

The remaining gauge-invariant variables are related to $h_{jk}^\text{TT}$ via Eqs.~\eqref{eq-sc-gi} and \eqref{eq-vc-gi}.
So although the scalars and the vector are generally nonzero, they are excited by the tensor modes.
There are indeed only two tensorial DoF's.
Now, let $\x^j$ and $\y^j$ are two orthonormal spacelike vectors, and $\hat\Omega\cdot \x=\hat\Omega\cdot \y=0$.
Define the polarization tensors \cite{Nishizawa:2009bf}
\begin{equation}
  \label{eq-def-pmpol}
  e^+_{jk}=\x_j \x_k-\y_j \y_k,\quad
  e^\times_{jk}=\x_j \y_k+ \y_j\x_k,
\end{equation}
for the $+$ and $\times$ polarizations, respectively.
Therefore, the scalar variables are
\begin{equation}
  \label{eq-s-sp}
  \Phi=\frac{1}{2}\Theta=-\frac{1}{2}(\cb s^+h_++\cb s^\times h_\times),
\end{equation}
where $\cb s^P=\cb s^{jk}e^P_{jk}/2$ and $h_P=h_{jk}^\text{TT}e_P^{jk}/2$.
This relation implies that the two tensor modes contribute to the scalar modes differently, in general.
One may say that the excitation of the scalar modes by the tensor ones is chiral, as the $+$ and $\times$ polarization tensors are linearily related to the left- and right-handed helicity basis in the following way,
\begin{gather}
  e^\text{L}_{jk}=\frac{1}{\sqrt 2}(e^+_{jk}-i e^\times_{jk}),\quad
  e^\text{R}_{jk}=\frac{1}{\sqrt 2}(e^+_{jk}+ie^\times_{jk}).
\end{gather}
If $\tilde s^+$ and $\tilde s^\times$ both vanish, neither of the scalar modes exists any longer.
Similarly, the vector modes depend on the tensor polarizations in the following way,
\begin{equation}
  \label{eq-s-vp}
  \Xi_j=-\left(\mathcal Xh_++\mathcal Yh_\times\right)\hat\vartheta_j-(\mathcal X h_\times-\mathcal Yh_+)\hat\varphi_j,
\end{equation}
where one defines 
\begin{equation}
  \label{eq-def-xy}
\mathcal X=-\brb s^j\x_j+\cb s^{jk}e^\text{x}_{jk}/2,\quad\mathcal Y=-\brb s^j\y_j+\cb s^{jk}e^\text{y}_{jk}/{2}.
\end{equation}
So the induction of the vector modes by the tensor modes is also chiral.
Moreover, it is possible that there may exist only the $\hat\vartheta_j$ component or the $\hat\varphi_j$ component, if $\mathcal Xh_++\mathcal Yh_\times=0$ or $\mathcal X h_\times-\mathcal Yh_+=0$, respectively.
Finally, if $\hat\Omega^\mu$ is an eigenvector of $\tb s^{\mu\nu}$, both the vector modes disappear.

The GW polarizations can be detected by interferometers, PTAs and Gaia mission.
Note that since all polarizations are excited by the two tensorial DoF's, the way the detectors respond to the polarizations is quite different from that in most of the theories studied before \cite{Will:2014kxa,Hou:2017bqj,Gong:2018ybk,Gong:2018cgj,Gong:2018vbo,Dong:2021jtd,Dong:2023bgt}.
That is, in the previous studies, different polarizations usually are generated by different DoF's, so the detector responses can be calculated separately for each DoF, and they are independent of each other.
Here, in SME, one shall consider the responses caused by the two tensorial DoF's.
So formally, the responses can be viewed as some linear combinations of the one for each polarization, treated as if it is an independent DoF.
In the following sections, the response functions of the detectors to the GW in SME will be discussed.

\section{Gravitational wave polarizations}
\label{sec-gw-pol}

Although this theory possesses two tensorial DoF's, there are more than two polarizations. 
To analyze the polarization content of the theory, one calculates the linearized geodesic deviation equation \cite{mtw,Hou:2017bqj},
\begin{equation}
  \label{eq-s-lin-gde}
  \frac{\ud^2x^j}{\ud t^2}=-R_{tjtk}x^k,
\end{equation}
where $x^j$ represents the deviation vector separating the adjacent test particles.
The electric component $R_{tjtk}$ of the Riemann tensor is \cite{Gong:2018cgj}
\begin{equation}
  \label{eq-s-rie}
  \begin{split}
  R_{tjtk}=&\frac{\omega^2}{2}h_{jk}^\text{TT}+\omega^2\hat\Omega_\mu\tb s^{\mu l}h_{l(j}^\text{TT}\hat \Omega_{k)}\\
  &-\frac{\omega^2}{4}(\delta_{jk}-\hat \Omega_j\hat \Omega_k)\cb s^{il}h_{il}^\text{TT}.
  \end{split}
\end{equation}
It is clear that the first term on the right-hand side represents the $+$ and $\times$ polarizations as in GR.
The physical meaning of the remaining parts can be seen by reexpressing them in terms of the following tensor basis \cite{Nishizawa:2009bf}
\begin{subequations}
  \label{eq-def-pols-ex}
\begin{gather}
  e^\text{x}_{jk}=\ho_j\x_k+\x_j\ho_k,\quad e^\text{y}_{jk}=\ho_j\y_k+\y_j\ho_k,\\
  e^\text{b}_{jk}=\x_j\x_k+\y_j\y_k,\quad e^\text{l}_{jk}=\sqrt{2}\ho_j\ho_k,
\end{gather}
\end{subequations}
which are simply the tensor basis for vector-$x$, vector-$y$, breathing (b) and longitudinal (l) polarizations, respectively.
It can be shown that 
\begin{equation}
  \label{eq-s-rie-1}
  \begin{split}
    R_{tjtk}=&\frac{\omega^2}{2}\Big[(h_+e^+_{jk}+h_\times e^\times_{jk})\\
    &- \left(\mathcal Xh_++\mathcal Yh_\times\right)e^\text{x}_{jk}-\left(\mathcal Xh_\times-\mathcal Yh_+\right)e^\text{y}_{jk}\\
  &-(\cb s^+h_++\cb s^\times h_\times)e^\text{b}_{jk}\Big].
  \end{split}
\end{equation}
The first line of the above equation corresponds to the first term in Eq.~\eqref{eq-s-rie}, and indeed, these are $+$ and $\times$ polarizations.
The second line of Eq.~\eqref{eq-s-rie-1} comes from the second term in Eq.~\eqref{eq-s-rie}, so there are vector-$x$ and vector-$y$ polarizations.
Finally, the last line of Eq.~\eqref{eq-s-rie-1} represents the breathing polarization, which comes from the second line in Eq.~\eqref{eq-s-rie}.
There is no longitudinal polarization.
All the extra polarizations are excited by the tensor modes, as clearly shown by the presence of $h_+$ and $h_\times$.
According to the discussion in the previous section, the excitation of the vector and the scalar modes is chiral, so the induction of the extra polarizations is also chiral.
When $\mathcal Xh_++\mathcal Yh_\times=0$, the vector-$x$ polarization disappears, while if $\mathcal Xh_\times-\mathcal Yh_+=0$, the vector-$y$ polarization disappears.
If $\hat\Omega^\mu$ is an eigenvector of $\tb s^{\mu\nu}$, both of the vector polarizations cease to exist, which is also true when $\mathcal X=\mathcal Y=0$.
Similarly, if $\tb s^+$ and $\tb s^\times$ both vanish, the breathing polarization is absent.

With Eq.~\eqref{eq-s-rie} or \eqref{eq-s-rie-1}, one can calculate the antenna pattern functions for the ground-based interferometers, given by \cite{Hou:2019wdg}
\begin{equation}
  \label{eq-s-drf}
  \mathcal F_P=F_P+\delta F_P.
\end{equation}
Here, $F_P=D^{jk}e^P_{jk}$ represents the standard response functions, and $D^{jk}=(\hat x^j\hat x^k-\hat y^j\hat y^k)/2$ is the detector configuration tensor for the two arms pointing in the directions $\hat x$ and $\hat y$ \cite{Poisson2014,Nishizawa:2009bf}.
The second term in Eq.~\eqref{eq-s-drf} is 
\begin{equation}
  \label{eq-def-dfp}
\delta F_P\equiv D^{jk}\left(2\hat\Omega_\mu\tb s^{\mu l}e_{lj}^P\hat\Omega_k-\frac{1}{2}e^\text{b}_{jk}\tb s^{il}e^P_{il}\right),
\end{equation}
which is the correction due to the presence of $\tb s^{\mu\nu}$. 
Note that here, $P=+,\times$, as there are only two tensorial DoF's.
Explicitly, one has
\begin{subequations}
  \label{eq-dfpm}
\begin{gather}
  \delta F_+=\mathcal XF_\text{x}-\mathcal YF_\text{y}-\cb s^+F_\text{b},\\
  \delta F_\times=\mathcal YF_\text{x}+\mathcal XF_\text{y}-\cb s^\times F_\text{b},
\end{gather}
\end{subequations}
where $F_\text{x}$, $F_\text{y}$ and $F_\text{b}$ are the standard antenna pattern functions for the vector-$x$, vector-$y$ and breathing polarizations, given by $F_{P'}=D^{jk}e^{P'}_{jk}$ with $P'=\text{x},\text{y},\text{b}$ \cite{Nishizawa:2009bf}.
Indeed, the ``physical'' antenna pattern functions $\mathcal F_P$ for $+$ and $\times$ polarizations are linear combinations of $F_P$'s and $F_{P'}$'s.
Note that the coefficients $\mathcal X$, $\mathcal Y$ and $\cb s^P$ are also functions of GW direction $\hat\Omega$ (even when $d=4$).
Since the extra polarizations are induced by $h_+$ and $h_\times$, it is better to define the effective polarization tensors,
\begin{subequations}
  \label{eq-def-eff-pol-0}
\begin{equation}
  \label{eq-def-eff-pol}
  E^P_{jk}=e^P_{jk}-\fre^P_{jk}-\cb s^Pe^\text{b}_{jk},
\end{equation}
for $P=+,\times$, and here,
\begin{equation}
  \label{eq-xy-rfs}
\mathfrak e^P_{jk}\equiv2\hat\Omega_{(j}e^P_{k)l}\tb s^{l\mu}\hat\Omega_\mu=\left\{
  \begin{array}{cc}
    \mathcal Xe^\text{x}_{jk}-\mathcal Ye^\text{y}_{jk}, & P=+,\\
    & \\
    \mathcal Ye^\text{x}_{jk}+\mathcal Xe^\text{y}_{jk}, & P=\times.
  \end{array}
\right.
\end{equation}
\end{subequations}
So one knows that 
\begin{equation}
  \label{eq-s-rie-2}
R_{tjtk}=\frac{\omega^2}{2}\sum_{P=+,\times}h_PE^P_{jk},
\end{equation}
which is useful for the later discussion.

To visualize $\delta F_P$ as functions of $\hat\Omega^j$, let us consider the case of $d=4$, and use the normal font for the tensor $\tb s^{\mu\nu}$ and its components.
So $\tilde s^{\mu\nu}$ (i.e., \textit{the} $\tb s^{\mu\nu}$ at $d=4$) contributes to $\delta F_P$.
Now, construct a coordinate system such that the two arms $\hat x$ and $\hat y$ are parallel to the coordinate axes.
As shown in previous expressions, in terms of the parameterization \eqref{eq-tis}, $\tilde s$ does not contribute to $\delta F_P$.
Now, one examines the contributions from $\mathring s^j$ and $\breve s^{jk}$, separately.
It is beneficial to further parameterize $\mathring s^j$ and $\breve s^{jk}$.
Since $\mathring s^j$ is a vector, it is natural to parameterize it in the following way,
\begin{equation}
  \label{eq-mrs}
\mathring s^j\equiv \mathring s(\sin o\cos\varsigma,\sin o\sin\varsigma,\cos o),
\end{equation}
with $\mathring s$ the magnitude of $\mathring s^j$.
By definition, $\breve s^{jk}$ is symmetric and traceless, so one may write it as a linear combination of five basic symmetric, traceless matrices
\begin{equation}
  \label{eq-def-mms}
  \breve s^{jk}=\sum_{n=1}^5\breve s^nM_n^{jk},
\end{equation}
where $M_1^{jk}=2\delta_1^{(j}\delta_2^{k)}$, $M_2^{jk}=2\delta^{(j}_1\delta^{k)}_3$, $M_3^{jk}=2\delta^{(j}_2\delta^{k)}_3$, $M_4^{jk}=\delta^j_1\delta^k_1-\delta^j_2\delta^k_2$, and $M_5^{jk}=\delta^j_1\delta^k_1-\delta^j_3\delta^k_3$.
Therefore, $\tilde s^{\mu\nu}$ is parameterized by nine constants: $\tilde s, \mathring s, o, \varsigma$, and $\breve s_n\,(n=1,2,3,4,5)$.
Since there are a lot of parameters in $\tilde s^{\mu\nu}$, it is better to set only one or two of them to nonzero values so as to clearly show their impacts on $\delta F_P$, and the various correlations in the next two sections.

\begin{figure}[h]
  \centering
  \includegraphics[width=0.45\textwidth]{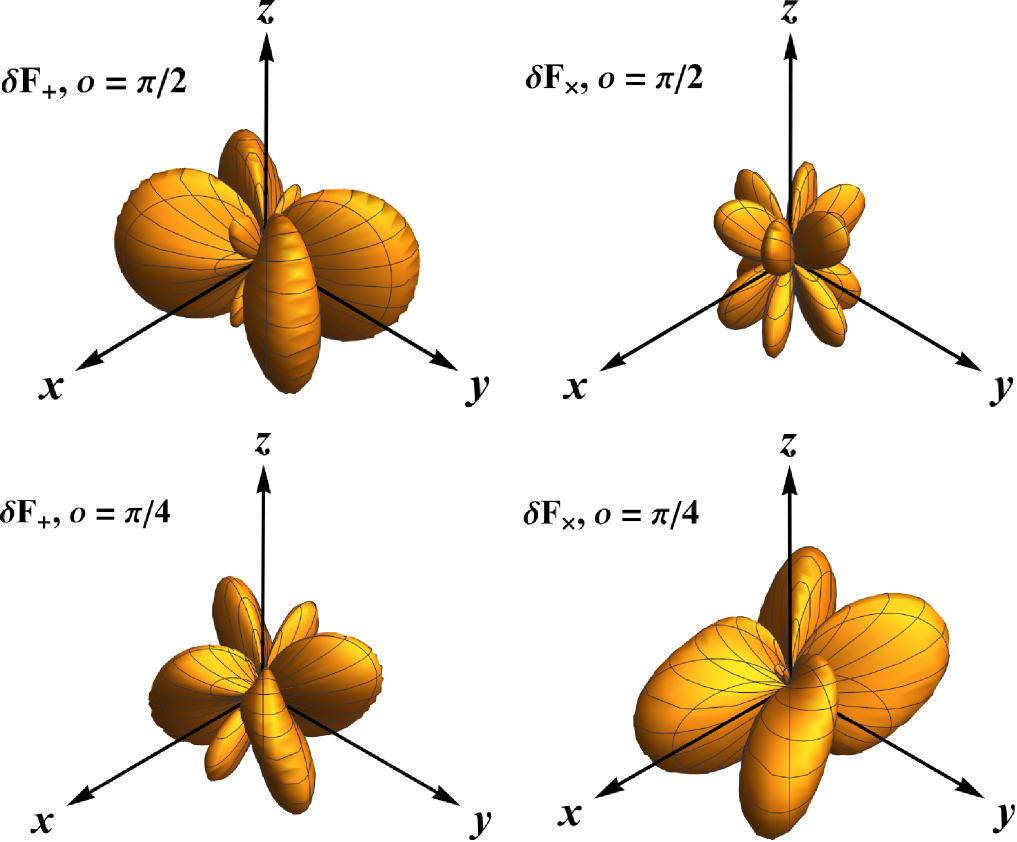}
  \caption{The corrections to the response functions due to $\hat v^j$.
  In all these plots, $\varsigma=0$. 
  }
  \label{fig-int-rps-vinx}
\end{figure}

First, consider the contribution of $\mathring s^j$, so  set $\breve s^{jk}=0$.
It is sufficient to choose $\mathring s=1$ for the purpose of demonstration.
Let us consider two cases.
In the first case, $o=\pi/2$ and $\varsigma=0$, i.e., $\mathring s^j$ is in the $+x$ direction, while in the second case, $o=\pi/4$ and $\varsigma=0$. 
$\delta F_P$ for these two cases are displayed in Fig.~\ref{fig-int-rps-vinx}.
It clearly shows that the responses due to $\mathring s^j$ are very different from the standard ones in GR, and the ones in some familiar modified theories of gravity \cite{Will:2014kxa,Nishizawa:2009bf}.
This is simply because $\delta F_P$'s are linear combinations of the vector polarizations by Eqs.~\eqref{eq-def-xy} and \eqref{eq-dfpm} for the case of $\mathring s^j$.

\begin{figure}[h]
  \centering
  \includegraphics[width=0.45\textwidth]{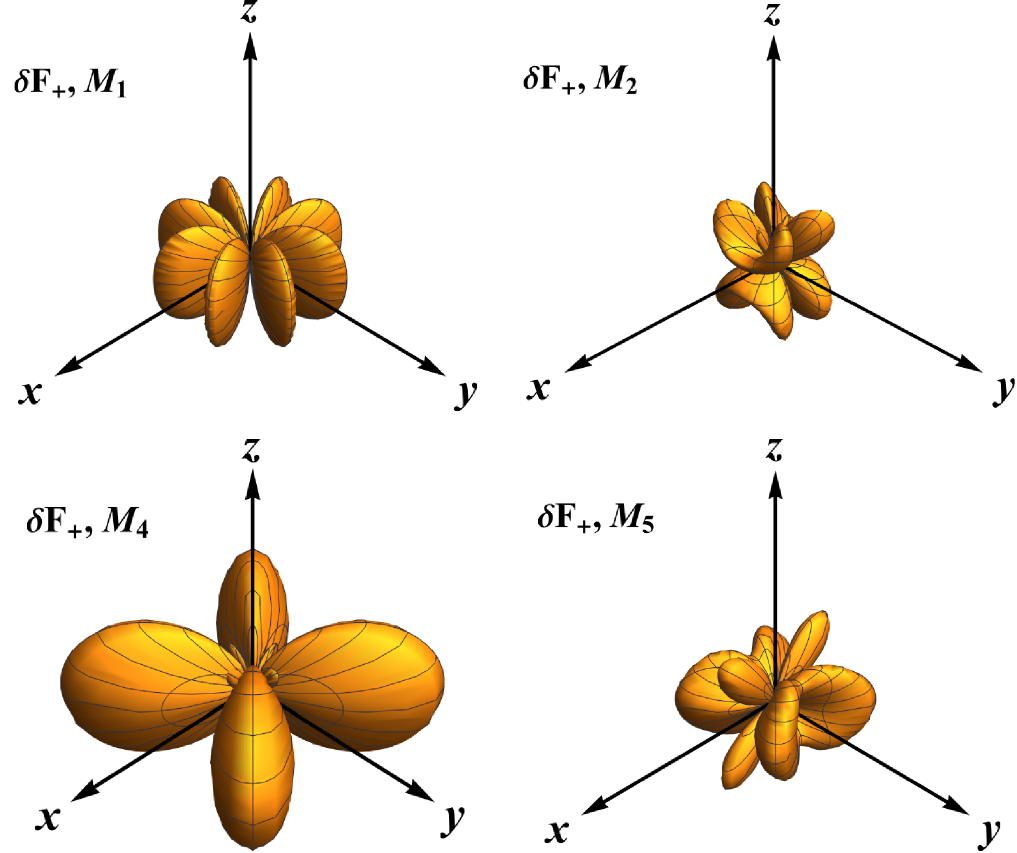}
  \caption{The corrections $\delta F_+$ to the response functions due to different components of $\breve s^{jk}$.
  The correction due to $M_3$ can be obtained by rotating the one for $M_2$ around the $z$-axis by $\pi/2$.}
  \label{fig-int-rps-ss}
\end{figure}

Second, switch off $\mathring s^j$ and turn on $\breve s^{jk}$.
It is also all right to set one of $\breve s_n$'s to 1 and the remaining to 0 to calculate $\delta F_P$, and then, iterate over the subscript $n$. 
In Fig.~\ref{fig-int-rps-ss}, $\delta F_+$ are shown for different types of $M$-matrices, and one can get $\delta F_+$ for $M_3$ by rotating the one for $M_2$ around the $z$-axis by $\pi/2$.
These plots, and those for $\delta F_\times$, are also different from the response functions for the standard polarizations \cite{Will:2014kxa,Nishizawa:2009bf}.
These corrections are linear combinations of the antenna pattern functions for the vector and the breathing polarizations.
Therefore, it is possible to measure the components of $\tilde s^{\mu\nu}$ based on the GW observations.

\section{Responses of pulsar timing arrays}
\label{sec-pta}

Pulsars are lighthouses in the universe.
They are rotating neutron stars or white dwarfs with strong magnetic fields.
In an ideal, empty space,  pulsars emit photons periodically, and millisecond pulsars are used as stable clocks \cite{Verbiest:2009kb}.
When there are perturbations to the space surrounding the pulsar and the earth, the received rate of the photon from the pulsar is altered.
The GW is such a kind of perturbation.
When the GW is present, the propagating time of the photon between the pulsar and the earth oscillates, which leads to the change in the time-of-arrival $T$. 
This change is called the timing residual $\mathcal R(T)$.
When the earth and pulsars are immersed in the SGWB, the timing residuals $\mathcal R_a(T)$ and $\mathcal R_b(T)$ of photons from two pulsars $a$ and $b$ are statistically correlated, and the correlation is given by a function $\mathcal C(\theta)=\langle\mathcal R_a(T)\mathcal R_b(T)\rangle$, where $\theta$ is the angle between the lines of sight to the pulsars, and the brackets $\langle\cdot\rangle$ mean to take the ensemble average over the SGWB.
The functional dependence of $\mathcal C$ on $\theta$ is related to the GW polarizations.
For example, $\mathcal C(\theta)$ in GR is given by the famous HD curve \cite{Hellings:1983fr}.
Thus, PTAs could also be used to detect the GW polarizations \cite{Lee2008ptac,Chamberlin:2011ev,Yunes:2013dva,Gair:2015hra,Hou:2017bqj,Gong:2018cgj,Gong:2018vbo}.
In this section, let us compute the responses of PTAs to the SGWB in SME.

In Section~\ref{sec-sme}, the gauge-invariant formalism was used to solve the EoM's.
Here, in order to compute $\mathcal R(T)$, one need explicitly determine the velocities of the photon, the earth and the pulsar.
So, one has to fix the gauge, e.g., imposing the following gauge fixing conditions,
\begin{equation}
  \label{eq-gfc}
h_{0\mu}=0,
\end{equation}
i.e., $\zeta_j=0$ and $\phi=\chi=0$.
This is consistent with Eq.~\eqref{eq-s-eom}.
Then, $\ddot\rho=-2\Phi$, and $\dot\varepsilon_j=-2\Xi_j$.
Let us start with the consideration of $\mathcal R(t)$ due to a monochromatic GW, propagating in $\hat\Omega$.
Then,
\begin{equation}
  \label{eq-hjk-gf}
  \begin{split}
  h_{jk}=&h_{jk}^\text{TT}+2\hat\Omega_{(j}\Xi_{k)}+2(\delta_{jk}-\hat\Omega_j\hat\Omega_k)\Phi\\
  =&\sum_{P=+,\times}h_PE^P_{jk}.
  \end{split}
\end{equation}
Comparing this with Eq.~\eqref{eq-s-rie-2} helps verify the correctness of this equation, as in the current gauge, $R_{tjtk}=-\ddot h_{jk}/2$.
It is further assumed that when there is no GW, the earth is at the origin of the coordinate system, and the pulsar is at $\vec x_p=L\hat n$, where $L$ is the distance between the earth and the pulsar, and $\hat n$ is the unit vector from the earth to the pulsar.
So their 4-velocities are $\bar u^\mu_\oplus=\bar u_p^\mu=\delta^\mu_0$.
At the same time, photon 4-velocity is $\bar u^\mu_\gamma=\gamma_0(1,-\hat n)$, where $\gamma_0=\ud t/\ud\lambda$ with $\lambda$ some arbitrary affine parameter. 
When the GW is present, the 4-velocities of the earth and the pulsar remain the same. 
This is because the geodesic equation for the pulsar takes the following form
\begin{equation}
  \label{eq-ge-p}
  \begin{split}
  0=&\frac{\ud u_p^\mu}{\ud \tau}+\Gamma^\mu{}_{\nu\rho}u_p^\nu u_p^\rho\\
  \approx& (u_p^0)^2\left(\frac{\ud u_p^\mu}{\ud t}+\Gamma^\mu{}_{00}\right)+u_p^0u_p^\mu\frac{\ud u_p^0}{\ud t},
  \end{split}
\end{equation}
where $\tau$ is the proper time, and $\Gamma^\mu{}_{00}=0$ in the chosen gauge,  then $u^0_pu^\mu_p$ is constant.
Since initially, $u_p^\mu|_{t=t_i}=\bar u_p^\mu=\delta^\mu_0$, one concludes that $u^\mu_p=\bar u_p^\mu=\delta^\mu_0$. 
One can also show that for the earth, $u^\mu_\oplus=\delta^\mu_0$ even when there is the GW.
However, the 4-velocity of the photon, $u^\mu_\gamma$, is perturbed by the GW. 
Assume the perturbed photon 4-velocity is $u^\mu_\gamma=\bar u^\mu_\gamma+V^\mu$, then the geodesic equation for the photon becomes
\begin{equation}\label{eq-phogeo}
  \begin{split}
  0=&\frac{\ud u^\mu_\gamma}{\ud \lambda}+\Gamma^\mu{}_{\rho\sigma}u^\rho_\gamma u^\sigma_\gamma\\
  \approx&\gamma_0\frac{\ud V^\mu}{\ud t}+\Gamma^\mu{}_{\rho\sigma}\bar u^\rho_\gamma \bar u^\sigma_\gamma.
  \end{split}
\end{equation}
From this, one can obtain
\begin{equation}
  \label{eq-pho-pv}
  \begin{split}
  V^\mu=&\frac{\gamma_0h_{jk}}{\omega+\vec\Omega\cdot\hat n}\left[(\omega+\vec\Omega\cdot\hat n)\hat n^j\delta^\mu_k-\frac{\hat n^j\hat n^k}{2}\Omega^\mu\right].
  \end{split}
\end{equation}
where $h_{jk}$ is given by Eq.~\eqref{eq-hjk-gf}, used here for simplicity.
So the observed photon frequencies on the earth ($f_\oplus=-u_{\oplus,\mu}u^\mu_\gamma$) and on the pulsar ($f_p=-u_{p,\mu}u^\mu_\gamma$) are no longer the same.
Then, the relative frequency shift, or the redshift, is given by
\begin{equation}
  \label{eq-rel-fs}
  \begin{split}
  z\equiv\frac{f_p-f_\oplus}{f_\oplus}=&\frac{1}{2}\frac{\hat n^j\hat n^k}{1+c_\text{gw}^{-1}\hat\Omega\cdot\hat n}(h_{jk}|_\oplus-h_{jk}|_p)\\
  =&\sum_{P=+,\times}\mathcal I^Ph_Pe^{-i\omega T}(1-e^{i\varpi }),
  \end{split}
\end{equation}
where $\varpi=L(\omega+\vec\Omega\cdot\hat n)$.
In addition, one defines
\begin{subequations}
  \label{eq-def-olre}
\begin{gather}
\mathcal I^P=I'^{P}-\cb s^P I'^{\text{b}}-\mathfrak I^P,\\
  \mathfrak{I}^P=\left\{
    \begin{array}{cc}
      \mathcal XI'^{\text{x}}-\mathcal YI'^{\text{y}}, & P=+,\\
      \mathcal YI'^{\text{x}}+\mathcal XI'^{\text{y}}, & P=\times,
    \end{array}
  \right.
\end{gather}
\end{subequations}  
where $I'^P$ takes the similar form as the overlap reduction function in GR  \cite{Hellings:1983fr,Lee2008ptac}, 
\begin{equation}
  \label{eq-def-ipm}
I'^{P}=\frac{\hat n^j\hat n^ke^P_{jk}}{2(1+c^{-1}_\text{gw}\hat\Omega\cdot\hat n)},
\end{equation}
differing only in the GW speed.
$I'^{P'}\,(P'=\text{x},\text{y},\text{b})$ in the above equations are also given by the similar expressions to Eq.~\eqref{eq-def-ipm} with $P$ replaced by $P'$.
Obviously, the overlap reduction functions $\mathcal I^P$ are still linear combinations of the individual ones $I'^P,I'^{P'}$, with the coefficients functions of $\hat\Omega$.
In terms of the parameterization \eqref{eq-tis}, one knows that $\tb s$ appears only in the denominators of the above equations, while $\brb s^j$ and $\cb s^{jk}$ appear both in the denominators and in the squared brackets.
For $d=4$, $c_\text{gw}$ is independent of the GW frequency $\omega$.
However, for $d>4$, the dispersion happens, and $\mathcal I^P$, $I'^P$ and $I'^{P'}$ are functions of  $\omega$.

Up to now, one finishes the computation of the relative frequency shift due to the presence of a monochromatic GW.
If there is the SGWB, one would have to consider the contributions to the redshift of all monochromatic GWs. 
Let us assume that for the SGWB, 
\begin{equation}
  \label{eq-def-sgwb}
  h_{jk}^\text{TT}(x)=\sum_{P=+,\times}\int_{-\infty}^{\infty}\frac{\ud\omega}{2\pi}\int\ud^2\hat\Omega e_{jk}^Ph_P(\Omega^\mu)e^{i\Omega\cdot x}.
\end{equation}
The timing residual is thus
\begin{equation}
  \label{eq-tres}
  \begin{split}
  \mathcal R(T)=&\sum_{P=+,\times}\int\frac{\ud\omega}{2\pi}\int\ud^2\hat\Omega\frac{i}{\omega}\mathcal I^Ph_P\\
  &\times(e^{-i\omega T}-1)(1-e^{i\varpi}).
  \end{split}
\end{equation}
Usually, one assumes that the stochastic GW background is isotropic, stationary, and unpolarized \cite{Lee2008ptac}. 
This assumption could also be made for SME, even if the propagation of the GW is anisotropic, as the sources of the GW could be randomly distributed.
Moreover, the most recent observations also highly constrained the anisotropy of SGWB \cite{NANOGrav:2023tcn}.
So one assumes the following ensemble average \cite{Lee2008ptac},
\begin{equation*}
  \langle h^{*}_P(\Omega)h_P(\Omega')\rangle=\delta(\omega -\omega ')\delta^{(2)}(\hat \Omega-\hat \Omega')\delta^{PP'}\frac{\pi|h_c^P(\omega)|^2}{4\omega}, \label{eq-def-s-h}
\end{equation*}
where $h_P^*$ means to take the complex conjugation, and $h_c^P$ is the characteristic strain amplitude.
Then, the cross-correction function $\mathcal C(\theta)$ between the timing residuals of photons from two pulsars located at $L_a\hat n_a$ and $L_b\hat n_b$ is
\begin{equation}
  \label{eq-xc}
  \begin{split}
  \mathcal C(\theta)=&\Re\{\langle\mathcal R^*_a(T)\mathcal R_b(T)\rangle\}\\
  =&\sum_{P=+,\times}\int\ud\omega\frac{|h^P_c(\omega)|^2}{8\pi\omega^3}\int\ud^2\hat\Omega\mathcal I_a^P\mathcal I_b^P\mathcal P,
  \end{split}
\end{equation}
where $\Re$ means to take the real part, and $\mathcal P=1-\cos\varpi_a-\cos\varpi_b+\cos(\varpi_a-\varpi_b)$.
The average over $T$ has been taken \cite{Lee2008ptac}.
$\theta$ is the angle between $\hat n_a$ and $\hat n_b$.
In the short-wavelength limit, $\omega L_a,\omega L_b\gg1$, so $\mathcal P\approx1$ for $\theta\ne0$, while for $\theta=0$, $\mathcal P\approx2$ \cite{Lee2008ptac}.
Although the overlap reduction function $\mathcal I^P$ are linear combinations of $I'^{P}$ and $I'^{P'}$, the correction $\mathcal C(\theta)$ is not the linear combination of $C^{P}(\theta)$ and $C^{P'}(\theta)$, which are defined similarly to $\mathcal C(\theta)$ with $\mathcal I^P$ in the integrand of Eq.~\eqref{eq-xc} replaced by $I'^P$ and $I'^{P'}$, respectively.
There are couplings among $C^P$ and $C^{P'}$.
This is different from what have been found in other modified theories of gravity previously, where  $\mathcal C(\theta)$ is indeed a linear combination of  $C^P$ and $C^{P'}$ \cite{Lee2008ptac,Lee:2010cg,Lee:2013awh,Gong:2018cgj,Gong:2018vbo}.

\begin{figure}[htbp]
  \centering
  \includegraphics[width=0.45\textwidth]{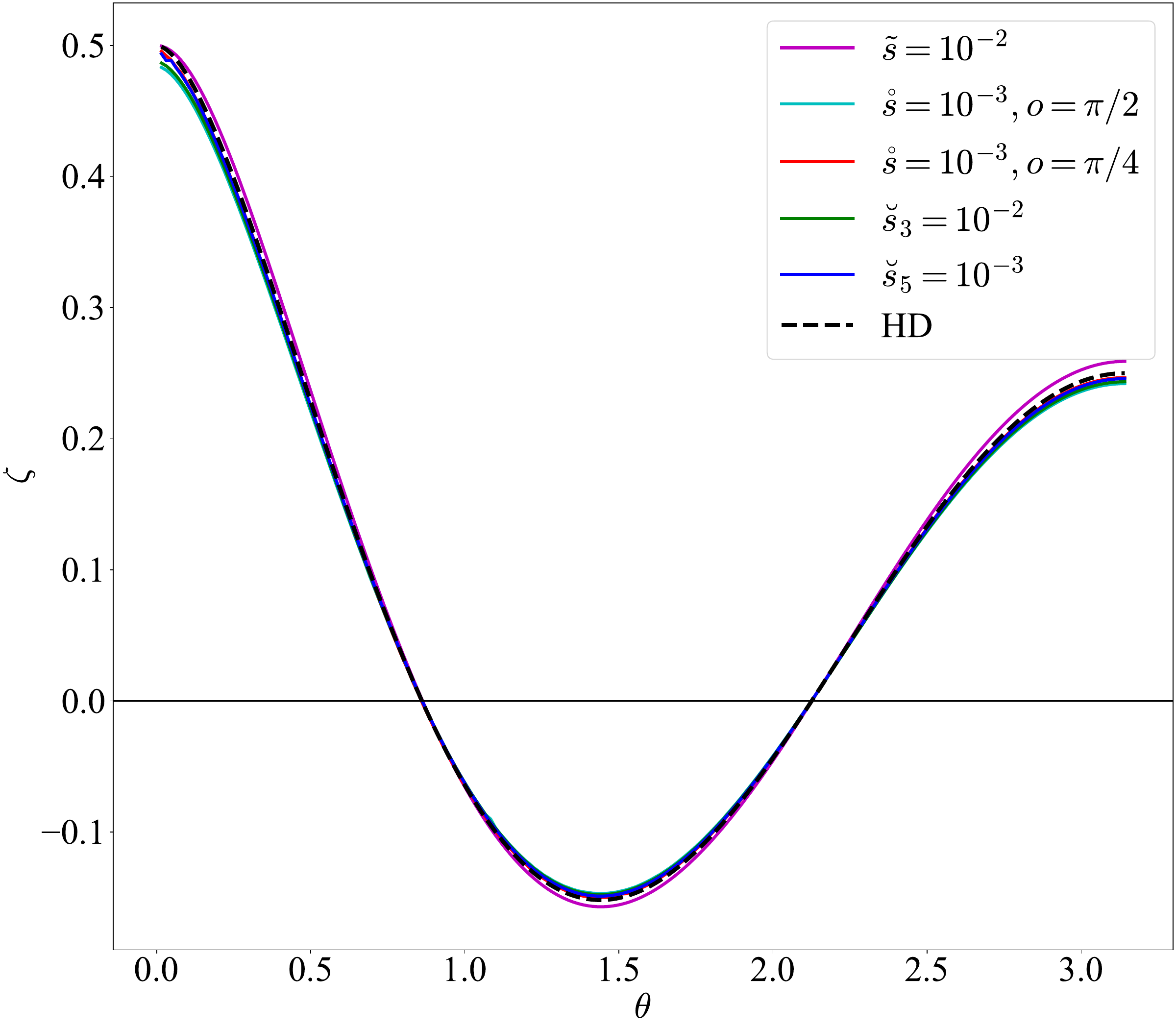}
  \caption{The normalized cross correlation function $\zeta(\theta)$.}
  \label{fig-zetas}
\end{figure}

To calculate the explicit dependence of $\mathcal C$ on $\theta$,  one can set 
\begin{equation}
  \label{eq-def-nab}
\hat n_a=(0,0,1),\quad \hat n_b=(\sin\theta,0,\cos\theta),
\end{equation}
without loss of generality. 
At the same time, parameterize $\hat\Omega=(\sin\theta_\text{gw}\cos\phi_\text{gw},\sin\theta_\text{gw}\sin\phi_\text{gw},\cos\theta_\text{gw})$.
One can compute $\mathcal C(\theta)$ as a function of $\theta$ after performing the integration \eqref{eq-xc}.
It is difficult to obtain the analytic expression, so the numeric integration shall be performed.
For the sake of being definitive, still consider the case of $d=4$.
Figure~\ref{fig-zetas} shows the normalized cross correlation functions $\zeta(\theta)=\mathcal C(\theta)/\mathcal C(0)$ for several choices of parameters.
Here, for simplicity, we will display $\zeta(\theta)$ by setting one or two of the components of $\tilde s^{\mu\nu}$ nonvanishing, while the remaining zero.
The magenta curve is for $\tilde s^{00}=\tilde s=\stvalue$ and the remaining components of $\tilde s^{\mu\nu}$ set to zero.
The brown and red curves are for $\mathring s=\mathringsv$ and $\varsigma=0$, but  $o=\ovalueI$ and $o=\ovalueII$, respectively.
The green and blue curves are for $\breve s_3=10^{-2}$ (one of the off-diagonal components of $\breve s^{jk}$) and $\breve s_5=10^{-3}$ (one of the diagonal components of $\breve s^{jk}$), respectively.
Finally, the black, dashed curve is the famous HD curve \cite{Hellings:1983fr}, given by
\begin{equation*}\label{eq-ncc-gr}
  \zeta(\theta)=\frac{3}{4}(1-\cos\theta)\ln\frac{1-\cos\theta}{2}+\frac{1}{2}-\frac{1-\cos\theta}{8}+\frac{\delta(\theta)}{2},
\end{equation*}
Here, $\delta(\theta)$ is the Dirac delta function, and the last term is nonzero when $\theta=0$, in which case, the photons come from the same pulsar, and $\zeta(0)$ is actually the (normalized) autocorrelation.
In principle, different components of $\tilde s^{\mu\nu}$ affect $\zeta(\theta)$ differently.
With the chosen values for the parameters, the magenta,  cyan, and green curves are different from the HD curve, but the remaining basically overlap with HD.
Given the recent observations by NANOGrav \cite{NANOGrav:2023gor}, PPTA \cite{Reardon:2023gzh}, EPTA+InPTA \cite{EPTA:2023sfo,EPTA:2023fyk} and CPTA \cite{Xu:2023wog}, one may conclude that it is currently difficult to use PTAs to distinguish GR and SME for the cases considered, here, which also happens to Einstein-\ae{}ther theory, for instance \cite{Gong:2018cgj}.
Fortunately, Gaia mission could also detect the SGWB.
The parameters that cannot be bounded strongly by PTAs might be restricted  by Gaia mission as shown in the next section.

One could also use the recently published data in Refs.~\cite{NANOGrav:2023gor,Reardon:2023gzh,EPTA:2023fyk,Xu:2023wog} to constrain the parameters  of $\tilde s^{\mu\nu}$.
This will be done in a follow-up work.
In the current paper, we focus on the theoretical predictions.

\section{Astrometric deflections}
\label{sec-gaia}

The presence of the GW not only modifies the measured frequency of a photon, but also changes the observed (angular) position of the pulsar.
This can be seen from the perturbation to the photon velocity \eqref{eq-pho-pv}, which contains nonvanishing spatial components, different from $\hat n$.
This leads to the change in the apparent astrometric position of a pulsar.
However, the astrometric position is not defined with respect to a coordinate system, but to the local inertia frame $\{e^\mu_{\hat a}\, (a=0,1,2,3)\}$ of an observer on earth. 
Here, $e^\mu_{\hat 0}=\delta^\mu_0$, no matter whether the GW exists or not \cite{Book:2010pf,Mihaylov:2018uqm}. 
The remaining basic vectors $e^\mu_{\hat j}$ vary due to the GW, so set $e^\mu_{\hat j}=\delta^\mu_j+\sfe^\mu_{j}$ with the second term the perturbation.
These spacelike vectors shall be parallel transported along the worldline of the observer, so they satisfy 
\begin{equation}\label{eq-patr}
\begin{split}
  0&=e^\nu_{\hat 0}\pd_\nu e^\mu_{\hat j}+\Gamma^\mu{}_{\rho\nu}e^\nu_{\hat0} e^\rho_{\hat j}\\
  &\approx\frac{\ud}{\ud t} \sfe^\mu_{j}+\Gamma^\mu{}_{0j},
\end{split}
\end{equation}
which implies 
\begin{equation}
  \label{eq-pte}
  \sfe^{\mu}_{j}=-\frac{1}{2}\left.\delta^\mu_kh_{jk}\right|_\oplus,
\end{equation}
evaluated at the earth.
Therefore, the observed astrometric position changes due to the deflection of the light trajectory and the rotation of the spatial tetrads.
It is given by $\hat n^{\hat j}+\delta n^{\hat j}=-u^\mu_\gamma e_\mu^{\hat j}/f_\oplus$, from which one has
\begin{equation}
  \label{eq-as-df}
  \delta\hat n^{\hat j}=\frac{1}{2}\left[(\hat n^j+c_\text{gw}^{-1}\hat\Omega^j)\frac{\hat n^k\hat n^lh_{kl}}{1+c_\text{gw}^{-1}\hat\Omega\cdot\hat n}-h_{jk}\hat n^k\right],
\end{equation}
evaluated at the earth, too.
Note that here, the short-wavelength approximation has been taken, so the so-called ``star terms'' have been dropped, leaving only the ``earth terms'' given above \cite{Book:2010pf,Mihaylov:2018uqm}.
Formally, Eq.~\eqref{eq-as-df} is similar to Eq.~(58) in Ref.~\cite{Book:2010pf} and Eq.~(20) in Ref.~\cite{Mihaylov:2018uqm}, where the speed of the GW was 1.
It also looks like the last term for tensor modes in Eq.~(62) in Ref.~\cite{Gong:2018vbo}, in which the GW speed can be superluminal.
$\delta\hat n^{\hat j}$ can also be rewritten as 
\begin{equation}
  \label{eq-as-df-2}
  \delta\hat n^{\hat j}=\sum_{P=+,\times}\mathcal J^{jP}h_P=\sum_{P=+,\times}\mathcal J^{jkl}E^P_{kl}h_P,
\end{equation}
where $\mathcal J^{jP}\equiv\mathcal J^{jkl}E^P_{kl}$, and following Ref.~\cite{Mihaylov:2018uqm}, one defines
\begin{equation}
  \label{eq-def-cj}
  \mathcal J^{jkl}=\frac{1}{2}\left[(\hat n^j+c_\text{gw}^{-1}\hat\Omega^j)\frac{\hat n^k\hat n^l}{1+c_\text{gw}^{-1}\hat\Omega\cdot\hat n}-\delta^{jl}\hat n^k\right].
\end{equation}
So $\mathcal J^{jP}$ measures the response of the astrometric position to the GW polarization $P$ with a unit amplitude.
Due to Eq.~\eqref{eq-def-eff-pol-0}, it is certainly a linear combination,
\begin{subequations}
  \label{eq-def-jsym}
\begin{gather}
  \label{eq-cj-lc}
  \mathcal J^{jP}=J'^{jP}-\cb s^P J'^{j\text{b}}-\mathfrak J^{jP},\\
  \mathfrak{J}^{jP}=\left\{
    \begin{array}{cc}
      \mathcal XJ'^{j\text{x}}-\mathcal YJ'^{j\text{y}}, & P=+,\\
      \mathcal YJ'^{j\text{x}}+\mathcal XJ'^{j\text{y}}, & P=\times,
    \end{array}
  \right.
\end{gather}
\end{subequations}
where $J'^{jP}$ and $J'^{jP'}$ are still defined just like $\mathcal J^{jP}$ with $E^P_{jk}$ replaced by $e^P_{jk}$ and $e^{P'}_{jk}$, respectively.

\subsection{Correlations between astrometric deflections}
\label{sec-ast-ast}

As the relative frequency shift \eqref{eq-rel-fs}, $\delta\hat n^{\hat j}$'s for different pulsars or stars are also statistically correlated. 
To represent the correlation between $\delta\hat n^{\hat j}_a$ and $\delta\hat n^{\hat j}_b$ of two pulsars $a$ and $b$, construct two sets of triads \cite{Mihaylov:2018uqm},
\begin{equation}
  \label{eq-triads-a}
  \hat n_a,\quad \hat u^I,\quad \hat u^J,
\end{equation}
for pulsar $a$, and
\begin{equation}
  \label{eq-triads-b}
  \hat n_b,\quad\hat u^\Upsilon,\quad\hat u^\Lambda,
\end{equation}
for pulsar $b$.
They satisfy \footnote{Here, we use different kind of superscripts from those in Ref.~\cite{Mihaylov:2018uqm}, because the superscripts $x$ and $y$ in their notation have been used to refer to the vector polarizations, and the superscripts $\theta$ and $\phi$ are used in Eqs.~\eqref{eq-def-pmpol} and \eqref{eq-def-pols-ex}.
To avoid confusions, it is better to use different labels.}
\begin{subequations}
  \label{eq-def-hus}
\begin{gather}
  \hat u^I={\mathcal N}(\hat n_a\times\hat n_b)\times\hat n_a,\quad \hat u^\Upsilon={\mathcal N}(\hat n_a\times\hat n_b)\times\hat n_b,\\
  \hat u^J=\hat u^\Lambda={\mathcal N}\hat n_a\times\hat n_b,
\end{gather}
\end{subequations}
where $\mathcal N=1/\sqrt{1-(\hat n_a\cdot\hat n_b)^2}$ is a normalization factor.
The correlations functions are defined to be
\begin{subequations}
\label{eq-def-as-cors}  
\begin{gather}
  \Gamma_{I\Upsilon}^P=\int\ud^2\hat\Omega\mathcal J_a^{IP}\mathcal J_b^{\Upsilon P},\quad
  \Gamma_{J\Lambda}^P=\int\ud^2\hat\Omega\mathcal J_a^{JP}\mathcal J_b^{\Lambda P},\label{eq-astast-1}\\
  \Gamma_{I\Lambda}^P=\int\ud^2\hat\Omega\mathcal J_a^{IP}\mathcal J_b^{\Lambda P},\quad
  \Gamma_{J\Upsilon}^P=\int\ud^2\hat\Omega\mathcal J_a^{JP}\mathcal J_b^{\Upsilon P},\label{eq-astast-2}
\end{gather}
\end{subequations}
where $\mathcal J^{LP}=\mathcal J^{jP}\hat u_{\hat j}^L$, and $\mathcal J^{\Sigma P}=\mathcal J^{jP}\hat u_{\hat j}^\Sigma$ with $L=I,J$ and $\Sigma=\Upsilon,\Lambda$.
In GR, it has been shown that the functions in Eq.~\eqref{eq-astast-2} vanish \cite{Mihaylov:2018uqm}.
Whether they are vanishing in SME needs to be examined.
Similar to $\mathcal C(\theta)$, the integrands of these correlations contain complicated coupling terms between $J'^{jP}$ and $J'^{jP'}$, by Eq.~\eqref{eq-def-jsym}.

\begin{figure}[htbp]
  \centering
  \includegraphics[width=0.43\textwidth]{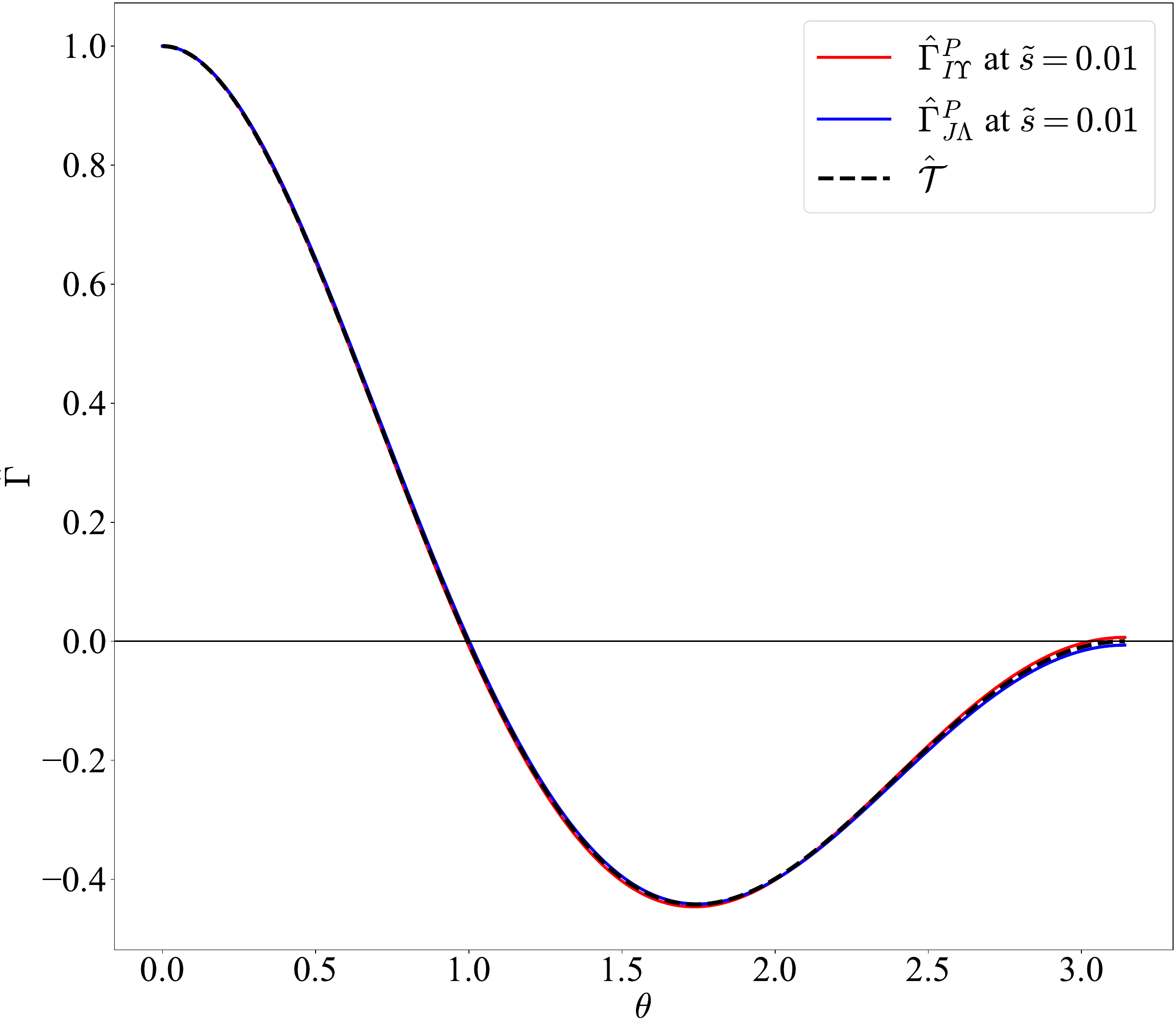}
  \caption{The normalized correlation functions of the astrometric deflections, $\hat\Gamma^P_{I\Upsilon}$ and $\hat\Gamma^P_{J\Lambda}$ ($P=+,\times$) for several choices of the parameters of $\tilde s^{\mu\nu}$.
  These are supposed to vanish in GR \cite{Mihaylov:2018uqm}.}
  \label{fig-ast-ast}
\end{figure}

To compute the explicit expressions for Eq.~\eqref{eq-def-as-cors}, one may still set $\hat n_a$ and $\hat n_b$ according to Eq.~\eqref{eq-def-nab}, and further assume
\begin{gather}
  \hat u^I=(1,0,0),\quad \hat u^J=(0,1,0),\\
  \hat u^\Upsilon=(\cos\theta,0,-\sin\theta),\quad\hat u^\Lambda=(0,1,0),
\end{gather}
consistent with Eq.~\eqref{eq-def-hus}.
Then, one could compute these correlations for $P=+,\times$, numerically.
For the case of $d=4$, the normalized correlations $\hat\Gamma^P_{I\Upsilon}$, $\hat\Gamma^P_{J\Lambda}$ at certain choices of the parameters of $\tilde s^{\mu\nu}$ are displayed in Fig.~\ref{fig-ast-ast}, with 
\begin{equation}
  \label{eq-def-normg}
  \hat\Gamma^P_{I\Upsilon}=\frac{\Gamma^P_{I\Upsilon}(\theta)}{\Gamma^P_{I\Upsilon}(0)},\quad \hat\Gamma^P_{J\Lambda}=\frac{\hat\Gamma^P_{J\Lambda}(\theta)}{\hat\Gamma^P_{J\Lambda}(0)}.
\end{equation}
In both panels, the black, dashed curve is the normalized $\hat{\mathcal T}=\mathcal T(\theta)/\mathcal T(0)$ in GR, in which \cite{Book:2010pf,Mihaylov:2018uqm},
\begin{equation}\label{eq-sc-gr}
  \begin{split}
  \mathcal T(\theta)=&\Gamma^P_{I\Upsilon}=\Gamma^P_{J\Lambda}\\
  =&\frac{2\pi}{3}-\frac{14\pi}{3}\sin^2\frac{\theta}{2}-\frac{8\pi\sin^4\frac{\theta}{2}}{\cos^2\frac{\theta}{2}}\ln\sin\frac{\theta}{2}.
  \end{split}
\end{equation}
In the upper panel, we plotted the correlations at $\tilde s=0.01$ with other parameters of $\tilde s^{\mu\nu}$ set to zero.
The red curve is for $\hat\Gamma^P_{I\Upsilon}$ and the blue for $\hat\Gamma^P_{J\Lambda}$.
As one can see, in GR, these two sets of correlation functions share the same curve (the dashed one), while in SME with $\tilde s$, their curves depart from the dashed one differently.
Nevertheless, these two correlation functions are independent of the polarizations $P$.
In fact, one could also draw the normalize cross correlations for the remaining choices of the parameters as in Fig.~\ref{fig-zetas}.
However, their curves almost overlap with GR's prediction.
In order to make the plots less busy, we only show the curves whose differences from $\hat{\mathcal T}$ are visible.

\begin{figure}[htbp]
  \centering
  \includegraphics[width=0.45\textwidth]{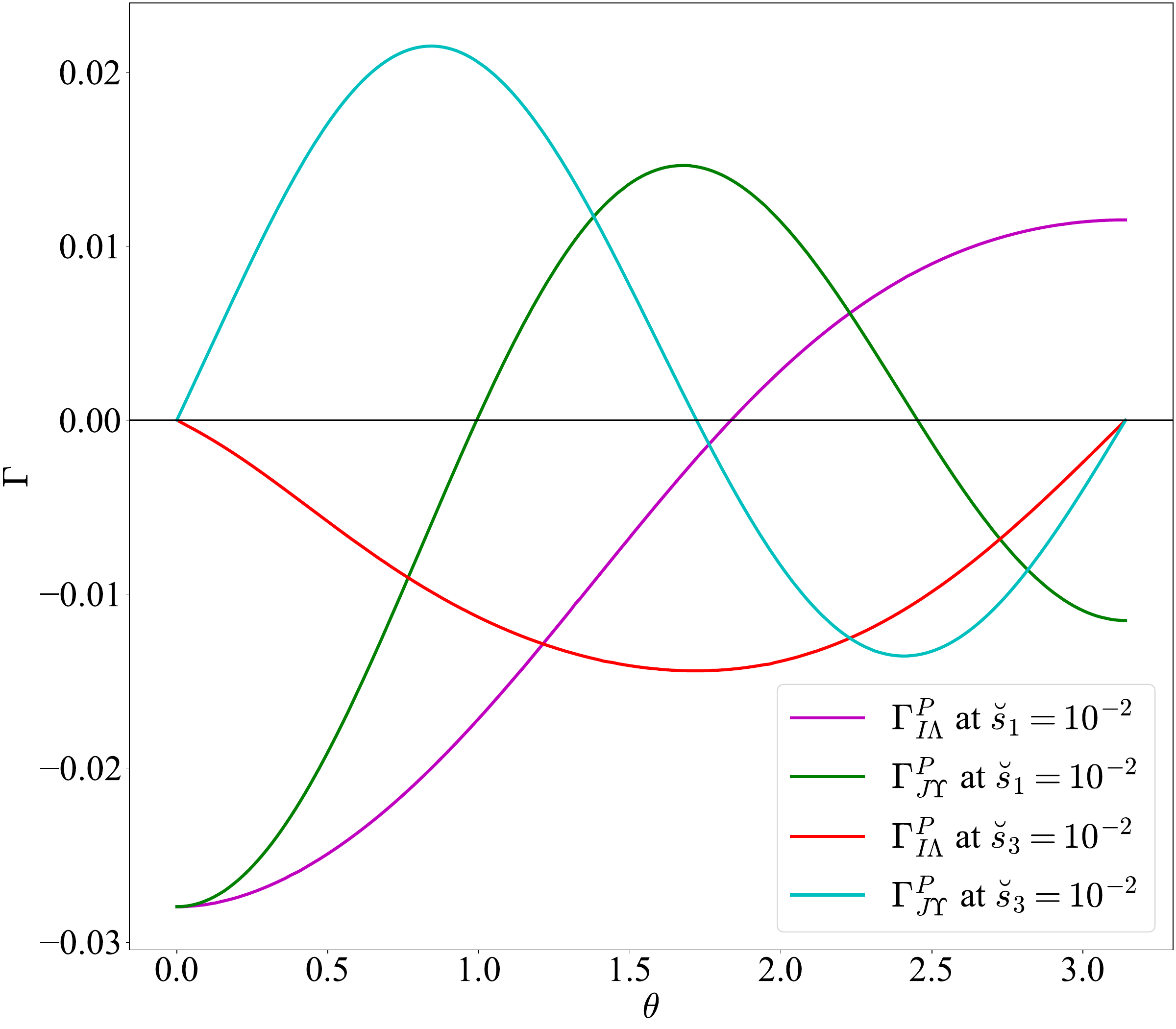}
  \caption{The normalized correlation functions of the astrometric deflections, $\hat\Gamma^P_{I\Lambda}$ and $\hat\Gamma^P_{J\Upsilon}$ ($P=+,\times$) for several choices of the parameters of $\tilde s^{\mu\nu}$.
  In GR, these functions are identically zero.}
  \label{fig-ast-ast-2}
\end{figure}

Although in GR, it was predicted that the correlations \eqref{eq-def-as-cors} vanish identically \cite{Mihaylov:2018uqm}, it is worth to check if they are zero in SME, too.
In Fig.~\ref{fig-ast-ast-2}, the correlations $\Gamma^P_{I\Lambda}$ and $\Gamma^P_{J\Upsilon}$ are drawn for some choices of the parameters of $\tilde s^{\mu\nu}$.
These correlations are zero in GR \cite{Book:2010pf,Mihaylov:2018uqm}, but in SME, they could be nonvanishing, as their expressions involve couplings between $J'^P$ and $J'^{P'}$.
As shown in Fig.~\ref{fig-ast-ast-2}, 
for the case of $\breve s_1=10^{-2}$ also considered in Fig.~\ref{fig-zetas},  $\Gamma^P_{I\Lambda}$ (magenta) and $\Gamma^P_{J\Upsilon}$ (green) are definitely different from zero.
The red and cyan curves are $\Gamma^P_{I\Lambda}$ and $\Gamma^P_{J\Upsilon}$ at $\breve s_3=10^{-2}$.
Note that none of these curves is normalized.
The extreme values of these curves are of the order of $10^{-2}$, consistent with $\breve s_1=\breve s_3=10^{-2}$.
When these parameters decrease to zero, these curves become the horizontal axis, reproducing GR's result.
The correlations $\Gamma^P_{I\Lambda}$ and $\Gamma^P_{J\Upsilon}$ for the remaining choices of parameters in Fig.~\ref{fig-zetas} are identically zero as in GR.
Therefore, the observation of such kind of correlations would be the smoking gun of the Lorentz violation in SME.

\subsection{Correlations between redshift and astrometric deflection}
\label{sec-red-ast}

Finally, the redshift \eqref{eq-rel-fs} is also correlated with the astrometric deflection \eqref{eq-as-df}.
The correlation functions can be defined to be \cite{Mihaylov:2018uqm}
\begin{equation}
  \Gamma^P_{z\Upsilon}(\theta)=\oint\ud^2\hat\Omega\mathcal I^P_a\mathcal J_b^{\Upsilon P},\quad \Gamma^P_{z\Lambda}(\theta)=\oint\ud^2\hat\Omega\mathcal I_a^P\mathcal J_b^{\Lambda P},
\end{equation}
where $\mathcal I_a^P$ is Eq.~\eqref{eq-def-olre} evaluated for the pulsar $a$.
In the case of GR, one has
\begin{gather}\label{eq-redast-gr}
  \mathcal Q(\theta)\equiv\Gamma^+_{z\Upsilon}(\theta)=\frac{4\pi}{3}\sin\theta+8\pi\sin^2\frac{\theta}{2}\tan\frac{\theta}{2}\ln\sin\frac{\theta}{2},\\
  \Gamma^\times_{z\Upsilon}=\Gamma^+_{z\Lambda}=\Gamma^\times_{z\Lambda}=0.
\end{gather}
Numerically compute the redshift-astrometric correlations in SME, and one obtains the following Fig.~\ref{fig-red-ast}, in which the so-called normalized correlation functions are displayed, 
\begin{equation}
  \label{eq-def-n-ra}
  \hat\Gamma^P_{z\Upsilon}=\frac{\Gamma^P_{z\Upsilon}(\theta)}{\max\{|\Gamma^P_{z\Upsilon}|\}}, \quad \hat\Gamma^P_{z\Lambda}=\frac{\Gamma^P_{z\Lambda}(\theta)}{\max\{|\Gamma^P_{z\Lambda}|\}}, 
\end{equation}
following Ref.~\cite{Mihaylov:2018uqm}.
In the above denominators, $\max$ means to take the maximum value of its argument.
Of course, these expressions are valid only when the maximum values are nonzero.
In Fig.~\ref{fig-red-ast}, the solid curves actually almost overlap with each other, which are $\hat\Gamma^P_{z\Upsilon}$'s for various choices of the parameters of $\tilde s^{\mu\nu}$ as displayed.
They are also almost identical to GR's prediction $\hat{\mathcal Q}$, which is represented by the black, dashed curve.
The dot-dashed curves are $\hat\Gamma^P_{z\Lambda}$ at $\breve{s}_1=10^{-2}$ (green) and $\breve{s}_3=10^{-2}$ (red).
Unlike in GR, this function is not identically zero at least for $\breve{s}_1$ and $\breve{s}_3$, but it is zero for the remaining choices of parameters except $\breve s_3$ listed in the legend box at the upper-right corner.
The extreme value of $\Gamma^P_{z\Lambda}$ (not normalized) for the dot-dashed green curve is about $-0.02$, and the one for the dot-dashed red is about $\pm0.04$.
If one decreases $\breve s_1$ and $\breve s_3$, $\Gamma^P_{z}$
\begin{figure}[htbp]
  \centering
  \includegraphics[width=0.44\textwidth]{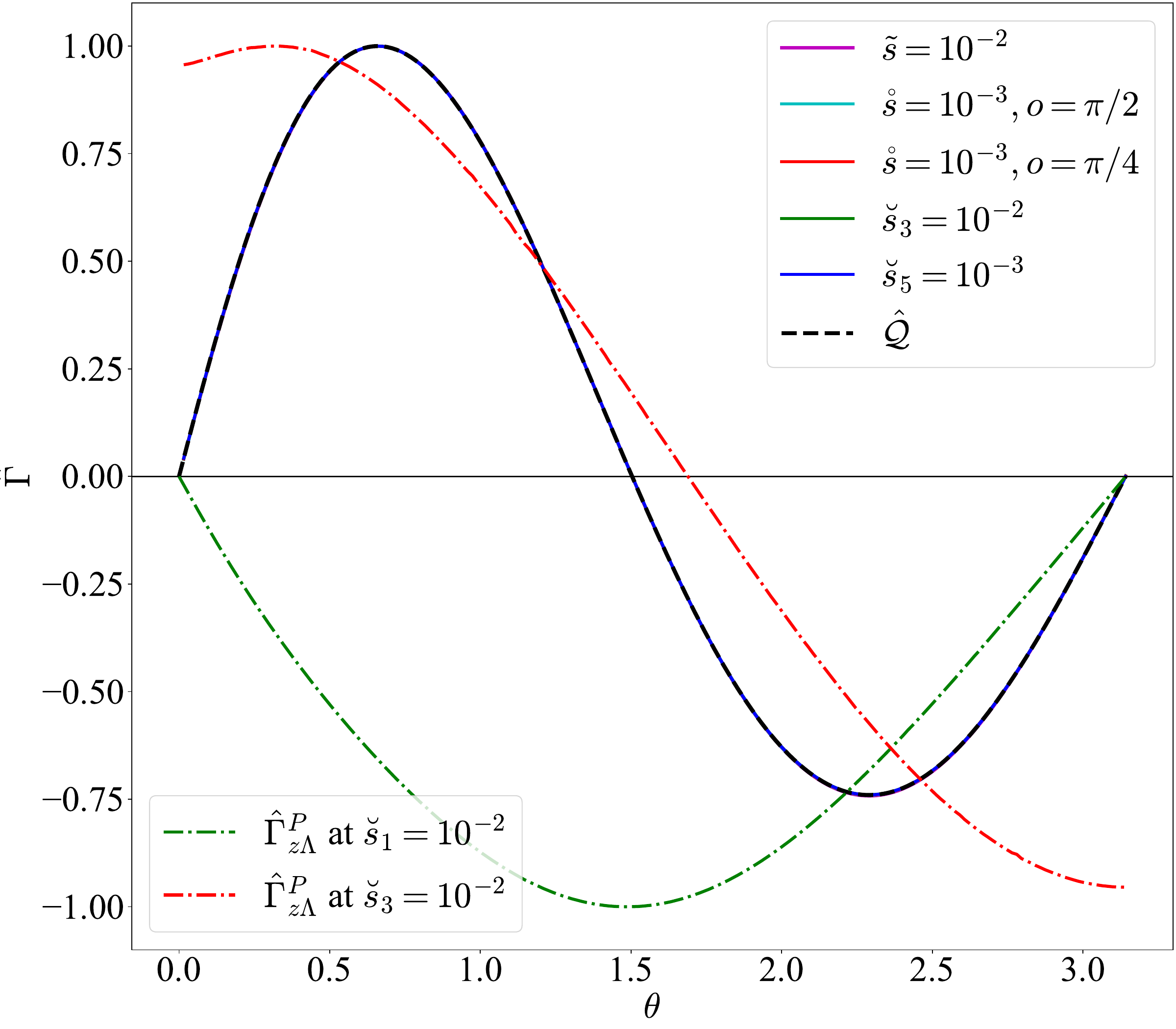}
  \caption{The normalized correlation functions of redshift and astrometric deflection, $\hat\Gamma^P_{z\Upsilon}$ and $\hat\Gamma^P_{z\Lambda}$ ($P=+,\times$) for several choices of the parameters of $\tilde s^{\mu\nu}$.}
  \label{fig-red-ast}
\end{figure}
This figure shows that although it may not be a good idea to use the correlation function $\Gamma^P_{z\Upsilon}$ to distinguish GR from SME, one may look for a nonvanishing $\Gamma^P_{z\Lambda}$.

\section{Conclusion}
\label{sec-con}

In this work, we studied the impact of the Lorentz violation introduced in the SME on the GW polarizations.
In the diffeomorphism-invariant sector with the $\tb s$ tensor, the SME systematically incorporates Lorentz-violating coefficients, which either define some preferred Lorentz frame, or provide special space directions.
Due to these coefficients, the tensor, vector, and scalar modes in this theory are coupled in such a manner that only two physical DoF's exist.
These DoF's are tensorial, and they excite all of the vector and scalar modes. 
They propagate at a speed other than one, depending on the propagation direction and thus, resulting in the anisotropic propagation.
Neither velocity nor amplitude birefringence could take place, and there is no dispersion effect.

Although there are two physical DoF's, five GW polarizations exist.
The extra vector-$x$, vector-$y$ and breathing polarizations are excited by the tensorial DoF's in a chiral way.
The dependence of the polarizations on the physical DoF's leads to the change in the detector responses to the GW.
The antenna pattern functions $\mathcal F_P$ for the two tensorial DoF's are now linear combinations of the standard interferometer responses $F'_{P}, F'_{P'}$ for polarizations, as if they were independent of each other.
The motion of photons immersed in the SGWB is also altered.
The total redshift $\mathcal I^P$ and the astrometric deflection $\mathcal J^{jP}$ also become some linear combinations of $I'^P,I'^{P'}$ and  $J'^{jP},J'^{jP'}$, respectively.
The changes in various correlation functions, redshift-redshift $\mathcal C(\theta)$, astrometric-astrometric $\Gamma^P_{L\Sigma}\; (L=I,J;\Sigma=\Upsilon,\Lambda)$, and redshift-astrometric $\Gamma^P_{z\Sigma}$, are more complicated, as their integrands are quadratic in $\mathcal I^P$ and $\mathcal J^{jP}$.
Therefore, these correlation functions include couplings among redshifts and astrometric deflections.

Numerical calculation helps visualize $\mathcal F_P$, $\mathcal C(\theta)$, $\Gamma^P_{L\Sigma}$ and $\Gamma_{z\Sigma}^P$.
One can clearly compare $\mathcal F_P$ and $F_P$ in Figs.~\ref{fig-int-rps-vinx} and \ref{fig-int-rps-ss}, justifying the use of interferometers to detect such kind of Lorentz violation.
Various cross correlation functions are shown in Figs.~\ref{fig-zetas}, \ref{fig-ast-ast}, \ref{fig-ast-ast-2} and \ref{fig-red-ast}.
There are certainly curves very similar to the standard ones, such as the HD curve $\zeta(\theta)$, $\mathcal T(\theta)$, and $\mathcal Q(\theta)$.
Very interestingly, correlations (e.g., $\Gamma^P_{I\Lambda},\Gamma^P_{J\Upsilon}$ and $\Gamma^P_{z\Lambda}$) might not be zero in SME, while they identically vanish in GR.
Observation of such correlations would be the smoking gun of SME.

In the current work, we considered mainly the theoretical aspects of the Lorentz violation brought about by the SME. 
In the follow-ups, we would like to constrain the theory based on the actual observational data from interferometers and PTAs.  
We would like also consider the effects of other Lorentz-violating operators, $\hb k$ and $\hb q$, on the GW polarizations. 
The induced birefringences by these operators might give more interesting phenomena and new constraints on the SME.

\begin{acknowledgements}
The authors were grateful for the discussion with Zhi-Chao Zhao.
This work was supported by the National Key Research and Development Program of China under Grant No.2020YFC2201503, the National Natural Science Foundation of China under grant Nos.~11633001 and 11920101003, and the Strategic Priority Research Program of the Chinese Academy of Sciences, grant No.~XDB23000000.
S. H. was supported by the National Natural Science Foundation of China under Grant No.~12205222, and by the Fundamental Research Funds for the Central Universities under Grant No.~2042022kf1062.
Tao Zhu was also supported by the National Natural Science Foundation of China under Grants No.12275238 and No. 11675143, the Zhejiang Provincial Natural Science Foundation of China under Grants No.LR21A050001 and No. LY20A050002, and the Fundamental Research Funds for the Provincial Universities of Zhejiang in China under Grant No. RF-A2019015. 
\end{acknowledgements}




%

\end{document}